\documentclass[letterpaper]{elsart}
\usepackage{lineno,hyperref}
\usepackage{amsmath}
\usepackage{amsfonts}
\usepackage{bm}
\usepackage{color}
\usepackage{colortbl}
\usepackage{dlfltxbcodetips}
\usepackage{multirow}
\usepackage{setspace}
\usepackage{subfigure}
\usepackage{verbatim}
\graphicspath{{./figures/}} 

\newcommand \D [2]{\frac{\partial #1}{\partial #2}}

\renewcommand{\vec}[1]{\bm{\mathrm{#1}}}
\newcommand{\tensor} [1] {\mathbb{#1}}

\def \tP{\tensor{P}}
\def \tF{\tensor{F}}
\def \tPdev{\tensor{P}_{\text{dev}}}

\def \cP{\mathcal{P}}
\def \tPdil{\tensor{P}_{\text{dil}}}

\def \F{\vec{F}}

\def \Fe{\F_{\text{e}}}

\def \U{\vec{U}}

\def \cF{\mathcal{F}}
\def \cG{\mathcal{G}}

\def \cT{\mathcal{T}}

\def \e{\vec{e}}

\def \f{\vec{f}}

\def \s{\vec{s}}
\def \u{\vec{u}}

\def \vsigma{\vec{\sigma}}

\def \x{\vec{x}}

\def \div{\nabla \cdot \mbox{}}
\def \grad{\nabla}
\def \lap{\nabla^2}

\def \Fe{\vec{F}^{\text{e}}}
\def \fe{\f^{\text{e}}}

\def \Ue{\vec{U}^{\text{e}}}
\def \Iqp{\text{I}qp}
\def \Eqp{\text{E}qp}
\def \varPsim{\varPsi_{\text{matrix}}}
\def \varPsif{\varPsi_{\text{fiber}}}

\def \fbi{\text{fb}i}
\def \fb{\text{fb}}
\def \tmo{\text{mucosa}}
\def \tcm{\text{CM}}
\def \tlm{\text{LM}}
\def \tf{\text{f}}
\def \ts{\text{s}}
\def \vsigma{\vec{\sigma}}
\def \te{\text{e}}
\begin{document}

\begin{frontmatter}



\title{A continuum mechanics-based musculo-mechanical model for esophageal transport}

\author{Wenjun Kou}
\address{Theoretical and Applied Mechanics, Northwestern
 University, 2145 Sheridan Road, Evanston, Illinois 60208, USA}

\author{Boyce E.~Griffith}
\address{Departments of Mathematics and Biomedical Engineering, University of North Carolina at Chapel Hill, Phillips Hall, Campus Box 3250, Chapel Hill, North Carolina 27599-3250, USA}

\author{John E.~Pandolfino}
\address{Department of Medicine, Feinberg School of Medicine, Northwestern University, 676 North Saint Clair Street, 14th Floor, Chicago, Illinois 60611, USA}
\author{ Peter J.~Kahrilas}
\address{Department of Medicine, Feinberg School of Medicine, Northwestern University, 676 North Saint Clair Street, 14th Floor, Chicago, Illinois 60611, USA}
\author{Neelesh A.~Patankar\corauthref{cor1}} \corauth[cor1]{Corresponding author.\ead{n-patankar@northwestern.edu}}
\address{Department of Mechanical Engineering, Northwestern
 University, 2145 Sheridan Road, Evanston, Illinois 60208, USA}

\begin{abstract}
In this work, we extend our previous esophageal transport model using an immersed boundary (IB) method with discrete fiber-based structural model, to one using a continuum mechanics-based model that is approximated based on finite elements (IB-FE). To deal with the leakage of flow when the Lagrangian mesh becomes coarser than the fluid mesh, we employ adaptive interaction quadrature points to deal with Lagrangian-Eulerian interaction equations based on a previous work (Griffith and Luo~\cite{Griffith2012IBFE}). In particular, we introduce a new anisotropic adaptive interaction quadrature rule.
The new rule permits us to vary the interaction quadrature points not only at each time-step and element but also at different orientations per element. This helps to avoid the leakage issue without sacrificing the computational efficiency and accuracy in dealing with the interaction equations. For the material model, we extend our previous fiber-based model to a continuum-based model. We present formulations for general fiber-reinforced material models in the IB-FE framework. The new material model can handle non-linear elasticity and fiber-matrix interactions, and thus permits us to consider more realistic material behavior of biological tissues. To validate our method, we first study a case in which a three-dimensional short tube is dilated. Results on the pressure-displacement relationship and the stress distribution matches very well with those obtained from the implicit FE method. We remark that in our IB-FE case, the three-dimensional tube undergoes a very large deformation and the Lagrangian mesh-size becomes about 6 times of Eulerian mesh-size in the circumferential orientation. To validate the performance of the method in handling fiber-matrix material models, we perform a second study on dilating a long fiber-reinforced tube. Errors are small when we compare numerical solutions with analytical solutions. The technique is then applied to the problem of esophageal transport. We use two fiber-reinforced models for the esophageal tissue: a bi-linear model and an exponential model. We present three cases on esophageal transport that differ in the material model and the muscle fiber architecture. The overall transport features are consistent with those observed from the previous model. We remark that the continuum-based model can handle more realistic and complicated material behavior. This is demonstrated in our third case where a spatially varying fiber architecture is included based on experimental study. We find that this unique muscle fiber architecture could generate a so-called pressure transition zone, which is a luminal pressure pattern that is of clinical interest. This suggests an important role of muscle fiber architecture in esophageal transport.
  
\end{abstract}


\end{frontmatter}

\section{Introduction}
The immersed boundary (IB) method 
was introduced to model blood flow through heart valves~\cite{Peskin1972}, and has been widely used to simulate biological fluid dynamics~\cite{Mittal&Iaccarino2005,liu2006immersed,Bhalla2013446,Griffith2012,kou2015fully}. The IB method has also been extended to deal with applications involving kinematic constraints~\cite{Bhalla2013446}, the electric field~\cite{Bhalla201488,liu2007immersed} and thermal fluctuations~\cite{atzberger2007stochastic}, among others. The IB method utilizes a Eulerian description of the momentum and continuity equations of the fluid-structure system, and a Lagrangian description of the displacements and stresses/forces of the structure (i.e. the solid). Thus it permits nonconforming discretizations of the fluid and the structure, and allows the Lagrangian mesh to move freely over the background Eulerian mesh. This approach does not require a body-fitted fluid mesh.
Variables on the two meshes communicate via integral transforms with delta function kernels. In the conventional IB method~\cite{CSPeskin02}, the Eulerian fluid flow is described on a regular Cartesian grid and the Lagrangian structure in the discretized form is modeled by families of elastic fibers that can generate stretching and bending forces. This material model is referred to here as the \textit{fiber-based material model}. The Lagrangian-Eulerian interaction equations are utilized to spread elastic forces from the Lagrangian nodes to the Eulerian (i.e. fluid) grid and interpolate the Eulerian fluid velocity to the Lagrangian nodes. This conventional IB method is referred to as the \textit{IB-fiber} method. The IB-fiber method is convenient to use in practice due to its simplicity in describing the elastic force. However, it also presents some challenges. First, the capability to model more realistic complex material behaviors is limited. The IB-fiber method uses a structure-based model (i.e. using beams, springs etc.); not a continuum-based model. Thus, it is difficult to incorporate a continuum-based description of material's elasticity that is used for many biological tissues. Second, there is the issue of fluid leaking through the immersed structure. The IB-fiber model uses Lagrangian nodes to interact with the fluid grid. The Lagrangian mesh size defined by neighboring Lagrangian nodes vary spatially and temporally as those nodes will move according to the dynamics. Leakage can occur if the Lagrangian mesh becomes much coarser than the Eulerian mesh~\cite{CSPeskin02}. Recently, Griffith and Luo~\cite{Griffith2012IBFE} introduced a version of the IB method that uses a finite element (FE) method to describe the elastic model. This method also allows interaction equations to be handled differently, such that the Lagrangian mesh can be coarser than the Eulerian grid without leakage. This new method here is referred to as the \textit{IB-FE} method. Here we extend our previous IB-fiber based esophageal transport model to a IB-FE based esophageal transport model.   

Esophageal transport is a bio-physical process that transfers food bolus from the pharynx to the stomach through a tube-like esophagus~\cite{Kahrilas1989}. It involves interactions between a liquid-like bolus, the esophageal wall, and neurally controlled muscle activation. The esophageal wall consists of mucosal, circular muscle, and longitudinal muscle layers. The muscle activation involves the activation of muscle fibers in, both, circular and longitudinal muscle layers~\cite{Kahrilas1997,Mittal2005}.  

In our previous work we developed a fully-resolved esophageal transport model based on the IB-fiber method. The model was first of its kind that integrated the bolus, the esophageal wall, and muscle activation together. It allowed us to investigate the underlying bio-physical mechanisms, such as roles of muscle activation and mucosa in normal and abnormal conditions~\cite{kou2015muscle}. However, the fiber-based material model is limited in characterizing more realistic elastic behavior of the multi-layered esophageal wall. Experiments show that the esophageal wall is generally characterized as an anisotropic nonlinear elastic material~\cite{Yang2006a,Yang2006b,Natali2009,Stavropoulou2009,Sokolis2013}, in particular, a so-called fiber-reinforced material~\cite{Limbert2002}. To better describe the material behavior of the esophageal tissue, we extend our fiber-based material model to fiber-reinforced continuum-based material model. We present a generic mathematical formulation to incorporate such fiber-reinforced material models into the IB-FE framework.

Esophageal transport problem is a numerically challenging problem. It involves multiple-length scales, from 0.3 mm to around 200 mm. It also involves very large deformations, which result from interactions among the fluid, structure, and muscle activation. As a result, leakage is a challenging issue. Our strategy in the previous IB-fiber-based model was to use a fine Lagrangian mesh in the inner-most layers, such that the Lagrangian mesh is always finer than the fluid mesh even under the largest dilation. However, such an approach requires to know the deformation level during the whole dynamics in advance. This may not be practical in many cases. It also adds to the computational cost. Here, we introduce a new approach to tackle the leakage issue following the work by Griffith and Luo~\cite{Griffith2012IBFE}. Specifically, Griffith and Luo~\cite{Griffith2012IBFE} introduced an \textit{adaptive interaction quadrature rule} to deal with interaction equations. Their adaptive interaction quadrature rule could vary time-step by time-step, element by element to avoid the leakage issue. But it could lead to a unnecessarily large number of interaction quadrature points and impact the computational efficiency in certain cases. In this work, we extend the previous adaptive interaction quadrature rule~\cite{Griffith2012IBFE}, referred to here as the \textit{isotropic adaptive interaction quadrature rule}, to a so-called \textit{anisotropic adaptive interaction quadrature rule}. The anisotropic adaptive interaction quadrature rule is able to handle the variation of the aspect ratio of Lagrangian elements, and permits us to vary interaction quadrature points orientation by orientation per element. Therefore, it not only helps to avoid the leakage issue even when the Lagrangian mesh becomes much coarser than the Eulerian mesh, it also reduces the number of interaction quadrature points at each time step. This is especially useful when a non-uniform Eulerian mesh is used or the Lagrangian structure undergoes a very large deformation. For our large-scale esophageal transport model presented here, cases using the anisotropic adaptive interaction quadrature rule run about twice faster than cases using the previous isotropic adaptive interaction quadrature rule.
 
The organization of this paper is as follows. The mathematical formulation is given in Section~\ref{part2_formulation}. In particular, the treatment of Lagrangian-Eulerian interaction equations and the formulation on continuum-based material models is discussed in detail. Two validation studies are presented in Section~\ref{part3_validation}. The IB-FE based esophageal transport model is described in Section~\ref{part4_model}. Three cases that differ in the material model and muscle fiber architecture are presented. Conclusions are presented in Section~\ref{part5_conclusion}.

\section{Mathematical formulation}
\label{part2_formulation}
\subsection{The immersed boundary method}
The IB-FE method~\cite{Griffith2012} employs an Eulerian description for the momentum equation and continuity equation, and a Lagrangian description for the deformation of the immersed structure and the resulting structural forces. 
Let $\x = (x_1, x_2, ...) \in \Omega  $ denote fixed Cartesian coordinates. $\Omega \subset \mathbb{R}^d, d$ = 2 or 3, denotes the fixed domain occupied by the entire fluid-structure system. We use $\s = (s_1, s_2,...) \in U $ to denote the Lagrangian coordinates attached to the immersed structure, where $U$ denotes the Lagrangian domain in the reference configuration. We let $\vec{\chi}(\s,t) \in \Omega$ denote the physical position of material point $\vec{s}$ at time $t$. We denote the physical region occupied by the structure and fluid at time $t$ as $\Omega^\ts(t) = \vec{\chi}(U,t) \subseteq \Omega$ and $\Omega^\tf(t) = \Omega \setminus \Omega^\ts(t)$, respectively. Since we consider here that the structure is immersed in the fluid, the fluid-structure interface can be denoted as $\partial \Omega^\ts(t)$. The boundary of the whole domain, $\Omega$ is denoted as $ \partial \Omega$. Then, the governing equations in the fluid domain, $ \Omega^\tf(t)$ are,
\begin{align}
\rho^{\tf}\left(\D{\u ^{\tf}}{t}(\x,t) + \u^{\tf}(\x,t) \cdot \grad \u^{\tf}(\x,t) \right) -\div \vsigma^{\tf} &=0,  \label{eqn_momentum_f1} \\
\div \u^{\tf}(\x,t) &= q^{\tf} (\x,t), \label{eqn_continuity_f1}
\end{align}
where $\rho^{\tf},\u^{\tf}$ is the fluid density and velocity, and $\vsigma^{\tf}$ is the fluid stress. $q^{\tf} (\x, t) $ is the fluid source, whose temporal-spatial distribution and strength is assumed to be known. $q^{\tf} (\x, t) $ is non-zero only in the fluid domain.

We consider the structure is incompressible. Thus, governing equations in the structure domain, $\Omega^\ts(t)$ are
\begin{align}
\rho^{\ts}\left(\D{\u^{\ts}}{t}(\x,t) + \u^{\ts}(\x,t) \cdot \grad \u^{\ts}(\x,t) \right) -\div \vsigma^{\ts} &=0,  \label{eqn_momentum_s1} \\
\div \u^{\ts}(\x,t) &= 0. \label{eqn_continuity_s1}
\end{align}
The interface conditions on the fluid-structure interface, $\partial \Omega^\ts(t)$ are
\begin{align}
 \vsigma^{\tf} \cdot \vec{n} &= \vsigma^{\ts} \cdot \vec{n}, \label{eqn_momentum_fs1} \\
 \u^{\tf} &=\u^{\ts},  \label{eqn_continuity_fs1} 
\end{align}
where $\vec{n}$ is the outward normal unit vector to $\partial \Omega^\ts(t)$, outward being away from the domain $\Omega^\ts(t)$ of the structure. 

On the boundary of the entire domain, $\partial \Omega$
\begin{align}
\u^{\tf} &= \u^{\tf}_{\partial \Omega},  \label{eqn_bc_whole1}
\end{align}
where $\u^{\tf}_{\partial \Omega}$ is the specified velocity on the boundary, $\partial \Omega$.  Eq. \eqref{eqn_bc_whole1} considers the Dirichlet boundary conditions, but the formulation can be easily extended to consider other boundary conditions. Notice that since we consider the structure is immersed in the fluid, $\partial \Omega$ is the boundary of the fluid domain only.  Eq. \eqref{eqn_bc_whole1} stays unchanged in the IB formulation, and thus omitted during our following discussions in this section.

The governing equations above can be recast into an alternate form that is appropriate for IB implementation. A derivation of the IB governing equations using the principle of virtual work using a weak form has been provided by Boffi et al.~\cite{boffi2008hyper}. In Appendix A we provide an alternate derivation based on the strong form that is pedagogically useful.

In our current work, we consider that the fluid-structure system possesses a uniform mass density $\rho $, i.e. $\rho^{\ts} =\rho^{\tf} =\rho $, and a uniform dynamic viscosity $\mu$. This simplification implies that the immersed structure is neutrally buoyant and viscoelastic rather than purely elastic. Then the resulting IB governing equations can be written as below,
\begin{align}
\rho\left(\D{\u}{t}(\x,t) + \u(\x,t) \cdot \grad \u(\x,t) \right) & = -\grad p(\x,t) + \mu \lap \u(\x,t) + \vec{f}^{\text{e}}, \label{eqn_momentum3} \\
\div \u(\x,t) &= q(\x,t), \label{eqn_continuity3} \\
\vec{f}^{\text{e}} &= \int_{\Omega^\ts(t)} \div \vec{\sigma}^{\text{e}}  \delta(\x - \vec{\chi}(\s,t)) d\vec{\chi}(\s,t) \nonumber \\ 
 &- \int_{\partial \Omega^\ts(t)} \vec{\sigma}^{\text{e}} \cdot \vec{n}  \delta(\x - \vec{\chi}(\s,t)) da(\vec{\chi}(\s,t)), \label{eqn_eularf3} \\   
 \vec{\sigma}^{\text{e}} &= \cF[\vec{\chi}(\cdot,t)],\label{eqn_eulars3}
\end{align}
where $q(\x,t)$ is a given fluid source, such that $q(\x,t)|_{\Omega^\tf(t)} =q^{\tf} (\x,t) $, and $q(\x,t)|_{\Omega^\ts(t)} = 0$. $\vec{\sigma}^{\te}$ is the elastic stress of the immersed structure in the current configuration, i.e. the Cauchy elastic stress. $\vec{f}^{\text{e}}$ is the Eulerian force density. $\delta(\x)$ is the $d$-dimensional delta function. Eq. \eqref{eqn_eulars3} is the elastic stress equation that depends on the material model of the structure.

In the IB-FE method, it is convenient to describe the elasticity of the structure with respect to the Lagrangian material coordinate system. Thus, we introduce the first Piola-Kirchhoff stress tensor that describes the current force with respect to the reference configuration. The first Piola-Kirchhoff stress tensor $\tensor{P}$ is defined such that
\begin{equation}
 \int_{\partial V} \tensor{P}^{\text{e}}\cdot \vec{N} dA(\s) = \int_{\partial \vec{\chi}(V,t)} \vec{\sigma}^{\text{e}} \cdot \vec{n} da(\x), \label{eqn_PK1_define} 
\end{equation}
for any smooth region $V \subset U$. $\vec{N} $ and $\vec{n} $ are the outward unit normal along $\partial V$ and $\vec{\chi}(V,t)$, respectively. Based on the divergence theorem, eq. \eqref{eqn_PK1_define} also implies,

\begin{equation}
 \int_{V} \div \tensor{P}^{\text{e}}  d\s = \int_{\vec{\chi}(V,t)} \div \vec{\sigma}^{\text{e}} d\x.  \label{eqn_PK1_derive}
\end{equation}

Substituting eqs. \eqref{eqn_PK1_define} and \eqref{eqn_PK1_derive} into eq. \eqref{eqn_eularf3}, we obtain
\begin{align}
\vec{f}^{\te} &= \int_{U} \grad_{\s} \cdot \tensor{P}^{\te} \delta(\x - \vec{\chi}(\s,t)) d\s  \nonumber \\ 
&- \int_{\partial U} \tensor{P}^{\te} \cdot \vec{N}  \delta(\x - \vec{\chi}(\s,t)) dA(\s), \label{eqn_eularf3_1} 
\end{align}
where $\grad_{\s} \cdot \tensor{P}^{\te}$ is referred to as the \textit{Lagrangian internal force density}, and $\tensor{P}^{\te} \cdot \vec{N}$ is referred to as the \textit{Lagrangian transmission force density}. Like in the previous work~\cite{Griffith2012IBFE}, we introduce the Lagrangian force density, denoted as $\vec{F}^{\te}$, as follows
\begin{align}
\vec{f}^{\te} &= \int_{U} \Fe(\s,t) \delta(\x - \vec{\chi}(\s,t))  d\s.  \label{eqn_spread3}
\end{align}
$\vec{F}^{\te}$ needs to include both the internal and transmission force density. This can be done by employing the weak form as below,
\begin{align}
  \int_{U} \Fe (\s,t) \cdot \vec{V}(\s) d\s &= \int_{U} (\grad_{\s} \cdot \tensor{P}^{\te}) \cdot \vec{V}(\s) d\s - \int_{\partial U} \tensor{P}^{\te}\cdot \vec{N} \cdot \vec{V}(\s)  dA(\s) \nonumber \\
  &=-\int_{U} \tensor{P}^{\text{e}}  \colon \grad_{\s}\vec{V}(\s) d\s,  \label{eqn_spread_filter3_1}
\end{align}
for any Lagrangian test function $\vec{V}(\s)$ defined on $U$.

Based on eqs. \eqref{eqn_spread3}-\eqref{eqn_spread_filter3_1}, we obtain
\begin{align}
 \rho\left(\D{\u}{t} + \u \cdot \grad \u \right) & = -\grad p + \mu \lap \u + \vec{f}^{\text{e}}, \label{eqn_momentum4} \\
 \div \u &= q, \label{eqn_continuity4_1}  \\
 \fe(\x,t) &=\int_{U} \Fe(\s,t) \delta(\x - \vec{\chi}(\s,t))  d\s,  \label{eqn_spread4}     \\
 \int_{U} \Fe (\s,t) \cdot \vec{V}(\s) d\s &=-\int_{U} \tensor{P}^{\text{e}}  \colon \grad_{\s}\vec{V}(\s) d\s, \forall \vec{V}(\s),  \label{eqn_spread_filter4}\\ 
 \frac{\partial \vec{\chi}}{\partial t}(\s,t) &= \int_{\Omega} \u(\x,t) \delta(\x - \vec{\chi}(\s,t)) d\x, \label{eqn_interpolate4} \\
 \tensor{P}^{\text{e}} &= \cG[\vec{\chi}(\cdot,t)], \label{eqn_stress4}
\end{align}
where $\fe(\x,t)$ and $\Fe(\s,t)$ are the Eulerian and Lagrangian elastic force densities. 
Eq. \eqref{eqn_interpolate4} is the velocity-interpolating equation, which determines the velocity field on the Lagrangian system based on the Eulerian velocity field. Eq. \eqref{eqn_stress4} is the elastic stress equation that computes $\tensor{P}^{\text{e}}$ based on the material model of the structure.

Notice that eq. \eqref{eqn_spread_filter4} is also a projection, which projects the Lagrangian force density $\Fe (\s,t)$ into the function space defined by $\vec{V}(\s)$. Thus, we introduce a similar projection on the Lagrangian velocity field. This is done as below,
\begin{align}
  \Ue(\s,t) &= \int_{\Omega} \u(\x,t) \delta(\x - \vec{\chi}) d\x, \label{eqn_interpolate4_1} \\
 \int_{U} \frac{\partial \vec{\chi}}{\partial t}(\s,t) \cdot \vec{V}(\s) d\s &=\int_{U} \Ue(\s,t) \cdot \vec{V}(\s) d\s, \forall \vec{V}(\s),  \label{eqn_interpolate_filter4}
\end{align}
where $\Ue(\s,t)$ is an intermediate Lagrangian velocity field. The final Lagrangian velocity field $\frac{\partial \vec{\chi}}{\partial t}(\s,t)$ is a projection of the intermediate Lagrangian velocity field into the function space defined by $\vec{V}(\s)$. 
Consequently, we obtain a new set of equations as below,
\begin{align}
 \rho\left(\D{\u}{t} + \u \cdot \grad \u \right) & = -\grad p + \mu \lap \u + \vec{f}^{\text{e}}, \label{eqn_momentum5} \\
 \div \u &= q, \label{eqn_continuity4}  \\
 \fe(\x,t) &=\int_{U} \Fe(\s,t) \delta(\x - \vec{\chi}(\s,t))  d\s,  \label{eqn_spread5}     \\
 \int_{U} \Fe (\s,t) \cdot \vec{V}(\s) d\s &=-\int_{U} \tensor{P}^{\text{e}}  \colon \grad_{\s}\vec{V}(\s) d\s, \forall \vec{V}(\s),  \label{eqn_spread_filter5}\\ 
 \Ue(\s,t) &= \int_{\Omega} \u(\x,t) \delta(\x - \vec{\chi}) d\x, \label{eqn_interpolate5} \\
 \int_{U} \frac{\partial \vec{\chi}}{\partial t}(\s,t) \cdot \vec{V}(\s) d\s &=\int_{U} \Ue(\s,t) \cdot \vec{V}(\s) d\s, \forall \vec{V}(\s),  \label{eqn_interpolate_filter5} \\
 \tensor{P}^{\text{e}} &= \cG[\vec{\chi}(\cdot,t)]. \label{eqn_stress5}
\end{align}
We use eqs. \eqref{eqn_momentum5} - \eqref{eqn_stress5} in the current work. 

The above formulation differs from that of the IB-fiber method mainly in two parts: the treatment of the Lagrangian-Eulerian interactions and the description of material elasticity. We proceed by briefly discussing these two parts, respectively. We refer to the prior work~\cite{Griffith2012IBFE} for details on other aspects, including the spatial discretization, temporal discretization, and implementation.

We first discuss our treatment of the Lagrangian-Eulerian interactions. Notice that we drop the subscript $\e$ when denoting variables on the Lagrangian structures in the following equations. 

\subsection{Lagrangian-Eulerian Interactions}
Eqs. \eqref{eqn_spread5} - \eqref{eqn_interpolate_filter5} are equations on the Lagrangian-Eulerian interactions. Since those equations involve the weak form, we first introduce FE-based Lagrangian discretization. Let $\cT_h = \cup_e U^e$ be a triangulation of $U$ composed of elements $U^e$, and $\{\phi_l(\s)\}_{l=1}^M$ denote the Lagrangian basis functions. Then the physical configuration of the Lagrangian structure, $\vec{\chi}(\s,t)$ is approximated by
\begin{align}
 \vec{\chi}_h(\s,t) = \sum_{l=1}^{M}\vec{\chi}_l(t)\phi_l(\s), \label{eqn_pos_shapefunction}
\end{align}
where $\vec{\chi}_l(t)$ is the time-dependent physical position of the Lagrangian node $\s_l$. To satisfy the discretized power identity, we also approximate the Lagrangian force density $\F(\s,t)$ using the same shape functions as below
\begin{align}
 \vec{F}_h(\s,t) = \sum_{l=1}^{M}\vec{F}_l(t)\phi_l(\s), \label{eqn_force_shapefunction}
\end{align}
where $\vec{F}_l(t)$ is the nodal value of the Lagrangian force density. If we restrict the test functions to be linear combinations of the Lagrangian basis function, then we can obtain semi-discretized equations on the Lagrangian-Eulerian interactions as below
\begin{align}
  \f(\x,t) &=\sum_e\int_{U_e} \vec{F}(\s,t) \delta(\x - \vec{\chi}(\s,t))  d\s   \nonumber \\
           &=\sum_e \sum_{l}\left (\int_{U_e} \phi_l(\s) \delta(\x - \vec{\chi}(\s,t)) d\s \right )\vec{F}_l(t), \nonumber \\
           \label{eqn_spread6}     \\
 \sum_e \sum_{l} \left (\int_{U_e} \phi_l(\s)\phi_m(\s) d\s \right ) \F_l(t) &=-\sum_e\int_{U_e} \tensor{P} \cdot \grad_{\s} \phi_m(\s) d\s, \label{eqn_spread_filter6}   \\
 \U(\s,t) &= \int_{\Omega} \u(\x,t) \delta(\x - \vec{\chi}(\s,t)) d\x, \label{eqn_interpolate6} \\
  \sum_e \sum_{l}\left (\int_{U_e} \phi_l(\s) \phi_m(\s) d\s \right )\frac{\partial \vec{\chi}_l}{\partial t}(t) &=\sum_e \int_{U_e} \U(\s,t) \phi_m(\s) d\s.  \label{eqn_interpolate_filter6} 
\end{align}
Notice that the above equations involve five integrals over the Lagrangian domain. We remark that we use two different quadrature rules in approximating these five integrals, for reasons that will be discussed later. Specifically, we refer to the first quadrature rule as the \textit{interaction quadrature rule}, and apply it to the right-hand integrals in eqs. \eqref{eqn_spread6} and \eqref{eqn_interpolate_filter6}; we refer to the second quadrature rule as the \textit{elasticity quadrature rule} and apply it to eq. \eqref{eqn_spread_filter6} and the left-hand integral in eq. \eqref{eqn_interpolate_filter6}. More clearly, we let $\Iqp$ and $\omega_\Iqp$ denote the interaction quadrature points and the corresponding weights in each element, and $\Eqp$ and $\omega_\Eqp$ denote elasticity quadrature points and the corresponding weights in each element. Then eqs. \eqref{eqn_spread6} -\eqref{eqn_interpolate_filter6} can be written as,
\begin{align}
  \f(\x,t) =\sum_e \sum_{\Iqp \in U_e} \sum_{l}\phi_l(\s_{\Iqp})\omega_{\Iqp} & \delta(\x - \vec{\chi}(\s_{\Iqp},t)) \vec{F}_l(t),       \label{eqn_spread7} \\   
\sum_e \sum_{\Eqp \in U_e} \sum_{l}  \phi_l(\s_{\Eqp})\phi_m(\s_{\Eqp})\omega_{\Eqp}  \F_l(t) &=-\sum_e \sum_{\Eqp \in U_e} \tensor{P}_{\Eqp} \cdot \grad_{\s} \phi_m(\s_{\Eqp})\omega_{\Eqp}, \label{eqn_spread_filter7} \\
 \U(\s_{\Iqp},t) &= \int_{\Omega} \u(\x,t) \delta(\x - \vec{\chi}(\s_{\Iqp},t)) d\x, \label{eqn_interpolate7} \\
\sum_e  \sum_{\Eqp \in U_e} \sum_{l} \phi_l(\s_{\Eqp}) \phi_m(\s_{\Eqp})\omega_{\Eqp} \frac{\partial \vec{\chi}_l}{\partial t}(t) &=\sum_e \sum_{\Iqp \in U_e} \U(\s_{\Iqp},t)  \phi_m(\s_{\Iqp})\omega_{\Iqp}.   \label{eqn_interpolate_filter7}
\end{align}
The reason we have a separate quadrature rule in dealing with Lagrangian-Eulerian interactions is because the interaction quadrature rule is associated with (at least) three issues. First, the distribution of quadrature points relates to the leakage issue. In the conventional IB method, fluid leakage occurs when the Lagrangian mesh is coarser than the fluid mesh. Unlike the conventional IB method, the current IB-FE method employs \textit{interaction quadrature points}, instead of the Lagrangian nodes, to spread the Lagrangian forcing to, and interpolate the velocity from, the Eulerian field. This can be seen from eqs. \eqref{eqn_spread7} and \eqref{eqn_interpolate7}. Thus, to avoid the leakage, we only need to ensure that the ``submesh" composed of the interaction quadrature points is not coarser than the fluid mesh. To address the temporal-spatial variation of the Lagrangian mesh size during the interaction, we only need to vary the distribution of quadrature points. Second, the choice of the quadrature rule relates to the accuracy of the approximation. Unlike eq. \eqref{eqn_spread_filter7}, the integrand of the right-hand integral of eq. \eqref{eqn_spread7} contains a non-rational function, i.e. the delta function. Traditional quadrature rules that are optimal for polynomial integrand may not be optimal in eq. \eqref{eqn_spread7}. Third, the number of quadrature points relates to the computational cost. Notice that adding one more interaction quadrature point will bring (at least) two more evaluations of the non-rational delta function at each time-step. Thus, for three-dimensional large-scale simulations, a large number of the interaction quadrature points per element could greatly impact the computational efficiency. 

To address the above three issues, we introduce an \textit{ anisotropic adaptive interaction quadrature rule}. This quadrature rule allows the number and the distribution of interaction quadrature points to vary time-step by time-step, element by element, and orientation by orientation per element. This is discussed as below. 

\subsubsection{Anisotropic adaptive interaction quadrature rule}
We restrict our finite elements as bi-linear finite elements (i.e. four-node quadrilateral elements) in two dimensions and tri-linear elements (i.e eight-node hexahedral elements) in three dimensions. We discuss our treatment in three-dimensional cases as an example. 

Consider an arbitrary eight-node hexahedral element, $e$. In the reference configuration, it has three orientations, denoted as $(\hat{s}_{i}(e),i=1,..,3)$ and four edges along each orientation, denoted as $(d\hat{s}_{i}^a (e),i=1,..,3; a=1,..,4)$. In the current configuration, the four edges of the element $e$ along each orientation become $(d\hat{l}_i^a (e,t),i=1,..,3; a=1,..,4)$. Based on the mapping $\vec{\chi}(\s,t)$, we can compute the maximum length of the four edges along each orientation, denoted by $(d\hat{l}_i^{\text{max}} (e,t),i=1,..,3)$. 

Our fluid grid is a Cartesian grid. If the finest fluid grid size is uniform, i.e. $h_x = h_y =h_z =h$, then we compute mesh-size ratios along each orientation, denoted as $(r_i(e,t),i=1,..,3)$,
\begin{equation}
 r_i(e,t) = \frac{d\hat{l}_i^{\text{max}}(e,t)}{ h} (i=1,..,3). \label{eqn_meshsize_ratio_uniform}
\end{equation}
Notice that if the finest fluid grid size is non-uniform, one can compute $h = \text{min}(h_x, h_y,h_z)$, and then compute $(r_i(e,t),i=1,..,3)$ based on eq. \eqref{eqn_meshsize_ratio_uniform}. However, another formulation can be obtained if we know in advance the alignment information of the edges, $(d\hat{l}_i^a (e,t),i=1,..,3; a=1,..,4)$. For example, in our esophageal model, the structure is a three-dimensional long tube described by the cylindrical coordinate system $\s = (R, \Theta, Z)$. Thus, the third orientation is almost always aligned with $z$-direction during the simulation. Therefore we can compute $(r_i(e,t),i=1,..,3)$ in another way, when we have a nonuniform fluid grid size, where $h_x = h_y =h \neq h_z$,
\begin{align}
 r_i(e,t) &= \frac{d\hat{l}_i^{\text{max}}(e,t)}{ h} (i=1,2), \label{eqn_meshsize_ratio_nonuniform1} \\
 r_3(e,t) &=\frac{d\hat{l}_3^{\text{max}}(e,t)}{ h_z}. \label{eqn_meshsize_ratio_nonuniform2}
\end{align}
The above procedure is different from the previous isotropic adaptive interaction quadrature rule~\cite{Griffith2012IBFE}, which computes $(r_i(e,t),i=1,..,3)$ as below,
\begin{equation}
r_i(e,t) = r(e,t) = \frac{\displaystyle\max_{j=1,..,3}{d\hat{l}_j^{\text{max}}(e,t)}}{\displaystyle\min_{k=x,y,z}{h_k}} (i=1,..,3). \label{eqn_isotropic_rule}
\end{equation}
To facilitate the calculation of the number of the quadrature points, we define the \textit{number density} as the number of quadrature points per fluid grid along each orientation, denoted here as $n_\text{nd}$. Then for element $e$ at time $t$, we compute the number of the interaction quadrature points along each orientation as $(\lceil {r_i(e,t) \times n_\text{nd}} \rceil,i=1,..,3)$. $\lceil x \rceil$ denotes the smallest positive integer that is bigger than or equal to $x$. The total number of quadrature points in element $e$ at time $t$ is $\prod_{i=1,..,3} \left ( \lceil {r_i(e,t)\times n_\text{nd}}\rceil \right )$. 

The above procedure effectively addresses the three issues associated with the interaction quadrature points. For the first issue ofe leakage, it is easily seen that as long as we require $n_\text{nd} \geq 1.0$ and the interaction quadrature points are almost evenly distributed along each orientation, it is guaranteed that the ``submesh" composed by the interaction quadrature points is \textit{always} finer than the fluid mesh. For the second issue on the accuracy, we compare different quadrature rules in approximating a one-dimensional integral with integrand as a product of a polynomial and the four-point delta function~\cite{CSPeskin02}. We find that the Gaussian quadrature rule with one (or two) points could yield a satisfactory accuracy when the polynomial is up to the first (or second) order. So we choose Gaussian quadrature rule as our interaction quadrature rule. For the third issue on computational efficiency, we see that the number of quadrature points increases as the cube of $n_\text{nd}$ in three dimensional simulations. Thus, in our three dimensional simulations, we choose $n_\text{nd} = 1.0$. 

To summarize, in our three dimensional simulations, we choose Gaussian quadrature rule and the \textit{number density}, $ n_\text{nd} $ to be 1.0 for the adaptive interaction quadrature rule.

The elasticity quadrature rule is used to deal with eq. \eqref{eqn_spread_filter7} and the left-hand integral of eq. \eqref{eqn_interpolate_filter7}. We adopt the traditional quadrature rule that is used in the FE-based solid mechanics. This will be discussed in the following section on our material model.

\subsection{Continuum-based material model}

Compared with the conventional IB-fiber model, one advantage of our current FE-based formulation is the capability to model more complicated and realistic material elasticity. In our current work, we restrict our elasticity model to be so-called incompressible fiber-reinforced material model. This model considers the material to be incompressible and consist of a continuous matrix and families of elastic fibers. This type of material model is often used in modeling biological tissues~\cite{Limbert2002}, including the esophageal wall~\cite{Yang2006a,Yang2006b,Natali2009}. Gao et al.~\cite{Gao2014} adopted the fiber-reinforced model introduced by Holzapfel and Ogden~\cite{Holzapfel3445} to study ventricular mechanics based on the IB-FE method. Here we show a more general formulation on the fiber-reinforced elastic model in the IB-FE framework. The formulation is suitable for various types of fiber-reinforced models, including two types of models that are used in our current work on esophageal transport.

\subsubsection{Fiber-reinforced elastic model} 
Here, we describe the fiber-reinforced elastic model within the framework of hyper-elasticity~\cite{Limbert2002}. We assume that the elastic potential of the material exists and is denoted by $\varPsi$. It can be split into two parts: the elastic potential of the matrix, denoted by $\varPsim$ and the elastic potential of the fibers, denoted by $\varPsif$. Therefore, we obtain
\begin{equation}
 \varPsi = \varPsim+ \varPsif.  \label{eqn_elastic_model_psi_total}
\end{equation}
Before we give the formulation of the elastic potential, we need to introduce strain measurements. Let $\tensor{F} = \frac{\partial\chi}{\partial \s}$ and $ \tensor{C} =\tensor{F}^{T}\tensor{F} $ denote the deformation gradient and the right Cauchy-Green tensor, respectively. Then the elastic potential of the matrix $\varPsim$ is assumed to be of the form
\begin{align}
\varPsim &= \varPsim (I_1, I_2), \label{eqn_elastic_model_psi_matrix} \\
I_1 = \text{tr}(\tensor{C});   I_2 &= 0.5[I_1^2 - \text{tr}(\tensor{C}\tensor{C})], \label{eqn_elastic_model_I12}
\end{align}
where $I_1$ and $I_2$ are the first and second principle invariants of $\tensor{C}$ that characterize isotropic deformations. This form is often used to model isotropic material elasticity.

For fibers, we permit the elastic model to have several families of fibers, labeled $\text{fb}i, i=1,2...$, overlapping with each other. $\varPsif$ is assumed to be the sum of the elastic potentials from all families,
\begin{equation}
 \varPsif = \sum_{i}\varPsi_{\text{fb}i}.   \label{eqn_elastic_model_psi_fiberall}
\end{equation}
Each family of fibers is associated with certain orientation in its reference configuration, denoted as $\vec{a}_{\text{fb}i}$. Then, similar to many other biological models~\cite{Limbert2002}, the elastic potential of each family, $\varPsi_{\text{fb}i}$ is assumed to be of the form
\begin{align}
\varPsi_{\text{fb}i}  &= \varPsi_{\text{fb}i} (I_{\text{fb}i}), \label{eqn_elastic_model_psi_fiberone} \\
I_{\text{fb}i} &= \tensor{C} \colon \tensor{A}_{\text{fb}i}, \label{eqn_elastic_model_Ifb}
\end{align}
where $\tensor{A}_{\text{fb}i}=\vec{a}_{\text{fb}i} \bigotimes \vec{a}_{\text{fb}i}$. $I_{\text{fb}i}$ is the strain invariant that characterizes the stretching of a fiber. Specifically, if the fiber's reference configuration is the stress-free configuration, then $I_{\text{fb}i}$ is the square of its stretch ratio. 
Thus the total elastic potential is of the form below,
\begin{equation}
  \varPsi(I_1, I_2,I_{\text{fb}1},I_{\text{fb}2},.. ) =  \varPsim(I_1, I_2) + \sum_{i}\varPsi_{\text{fb}i} (I_{\text{fb}i}).  \label{eqn_elastic_model_psi_two}
\end{equation}
We can then compute an intermediate first Piola-Kirchhoff stress tensor, denoted by $\hat{\tP}$,
\begin{equation}
\hat{\tP} = \D {\varPsi}{\tF} = \sum_{k=1,2} \D{\varPsim}{I_k} \D{I_k}{\tF} + \sum_{i} \D{\varPsi_{\fbi}}{I_{\fbi}} \D{I_{\fbi}}{\tF}.   \label{eqn_elastic_model_PK1}  
\end{equation}
Notice that, 
\begin{equation}
 \D{I_1}{\tF} = 2\tF,   \D{I_2}{\tF} = 2(I_1\tensor{F}- 2\tensor{F}\tensor{C}), \D{I_{\fbi}}{\tF} = 2\tensor{F} \tensor{A}_{\fbi}. \label{eqn_elastic_model_dpsi_12}
\end{equation}
Thus, we get a simplified equation for $\hat{\tP}$ below,
\begin{equation}
 \hat{\tP} =  2\D{\varPsim}{I_1} \tensor{F} + 2\D{\varPsim}{I_2} (I_1\tensor{F}- 2\tensor{F}\tensor{C} ) + 2 \sum_{i}\D{\varPsi_{\fbi}}{I_{\fbi}} \tensor{F} \tensor{A}_{\fbi}. \label{eqn_elastic_model_pk1_2}
\end{equation}
Notice that the immersed structure is assumed to be incompressible, and the isotropic stress (i.e. the negative hydrodynamic pressure) acts as a Lagrange multiplier to enforce incompressibility. We find that it is important to only keep the deviatoric part of $\hat{\tP}$ in eq. \eqref{eqn_elastic_model_pk1_2}. Otherwise, we could have non-zero elastic stress even when no deformation occurs (i.e. $ \hat{\tP} \neq 0$, even when $\tensor{F}$  is the identity tensor). So we compute the deviatoric component of the intermediate first Piola-Kirchhoff stress tensor, denoted by $\tPdev$ below,
\begin{equation}
 \tPdev = \cP (\hat{\tP})=  \hat{\tP}- 1/3( \hat{\tP} \colon \tensor{F})\hat{\tP} ^{-T}. \label{eqn_elastic_model_pk1_dev}
\end{equation}
Moreover, the incompressibility condition is only solved in the Eulerian description, and it does not guarantee that the volume conservation holds in the Lagrangian description after velocity interpolation. Therefore, we add n \textit{dilatational stress} component as penalty for volume changes in Lagrangian finite elements. This stress is denoted as $\tPdil$ and computed as below,
\begin{equation}
 \tPdil = 2 \beta_s J(J-1) \tensor{F}^{-T}, \label{eqn_elastic_model_pk1_dil}
\end{equation}
where $J=$det$\tensor{F}$ and $\beta_s$ is a penalty parameter.
Then, we compute the first Piola-Kirchhoff stress tensor $\tP$ in eq. \eqref{eqn_stress5} as below,
\begin{equation}
 \tP =\tPdev + \tPdil.  \label{eqn_elastic_model_pk1_final}
\end{equation}
In our implementation, we use the third order Gaussian quadrature rule for the deviatoric stress $\tPdev$ and first order Gaussian quadrature rule for the dilatational stress $\tPdil$ to avoid volumetric locking. We also use third order Gaussian quadrature rules to approximate the left-hand sides in eqs. \eqref{eqn_spread_filter7} and \eqref{eqn_interpolate_filter7}.

\section{Validation Studies}
\label{part3_validation}
\subsection{ Validation of the capability to handle large deformation without leakage: dilation of a short tube}
 
We present three-dimensional test cases that are relevant to the problem of esophageal transport. Other test cases relevant to the IB-FE method can be found in prior work (\cite{Griffith2012IBFE}, \cite{Gao2014}). Our first three-dimensional case is dilation of a short tube to validate the capabilities of our method in dealing with very large deformation. 
In this case, a cylindrical tube is immersed in a fluid box. The tube is described in the cylindrical coordinates $\s = (R, \Theta, Z)$, with $0.5 \leq R \leq 1, 0 \leq \Theta \leq 2\pi, 0 \leq Z \leq 1$ in the initially stress-free configuration. The top and bottom of the tube is fixed through the penalty method. The tube is assumed to be isotropic neo-Hooken material, with an elastic model 
\begin{equation}
 \varPsi = \varPsim(I_1) = \frac{C_1}{2} (I_1 -3), \label{eqn_validation_case1_psi}
\end{equation}
where $C_1$ is the modulus and non-dimensionalized to be $C_1 = 1.0$. Thus, based on eq. \eqref{eqn_elastic_model_pk1_2}, we obtain $\hat \tP = C_1 \tensor{F}$. The penalty parameter, $\beta_s$ in eq. \eqref{eqn_elastic_model_pk1_dil} is set to be $10C_1$. 

The fluid is described in Cartesian coordinates $\x = (x, y, z)$, with $-2.0 \leq x \leq 2.0, -2.0 \leq y \leq 2.0, -0.0 \leq z \leq 1.0$. The density and viscosity of the fluid is set to be 1.0. Here, we also specify a fluid source $q(\x, t)$ in eq. \eqref{eqn_continuity4} in order to dilate the tube,
\begin{equation}
 q(\x,t) = \begin{cases} q_0 (1.0 - e^{-t}) & \mbox{if } (x^2 + y^2 > 0.4^2, z \in [0.1,0.9]), \\
 0 & \mbox{otherwise}, \end{cases} \label{eqn_validation_case1_q} 
\end{equation}
where $q_0 = 0.015625$ in our current study. 

The tube is discretized using eight-node hexahedral finite elements. The number of elements along $(R, \Theta, Z)$ orientation is $(n_R, n_\Theta, n_Z) = (5, 20 ,50)$. The fluid mesh sizes in x, y, and z directions are $h_x = h_y = h_z = 0.5$. 

\begin{figure}[ht] 
 \centering
 \includegraphics[scale = 0.4]{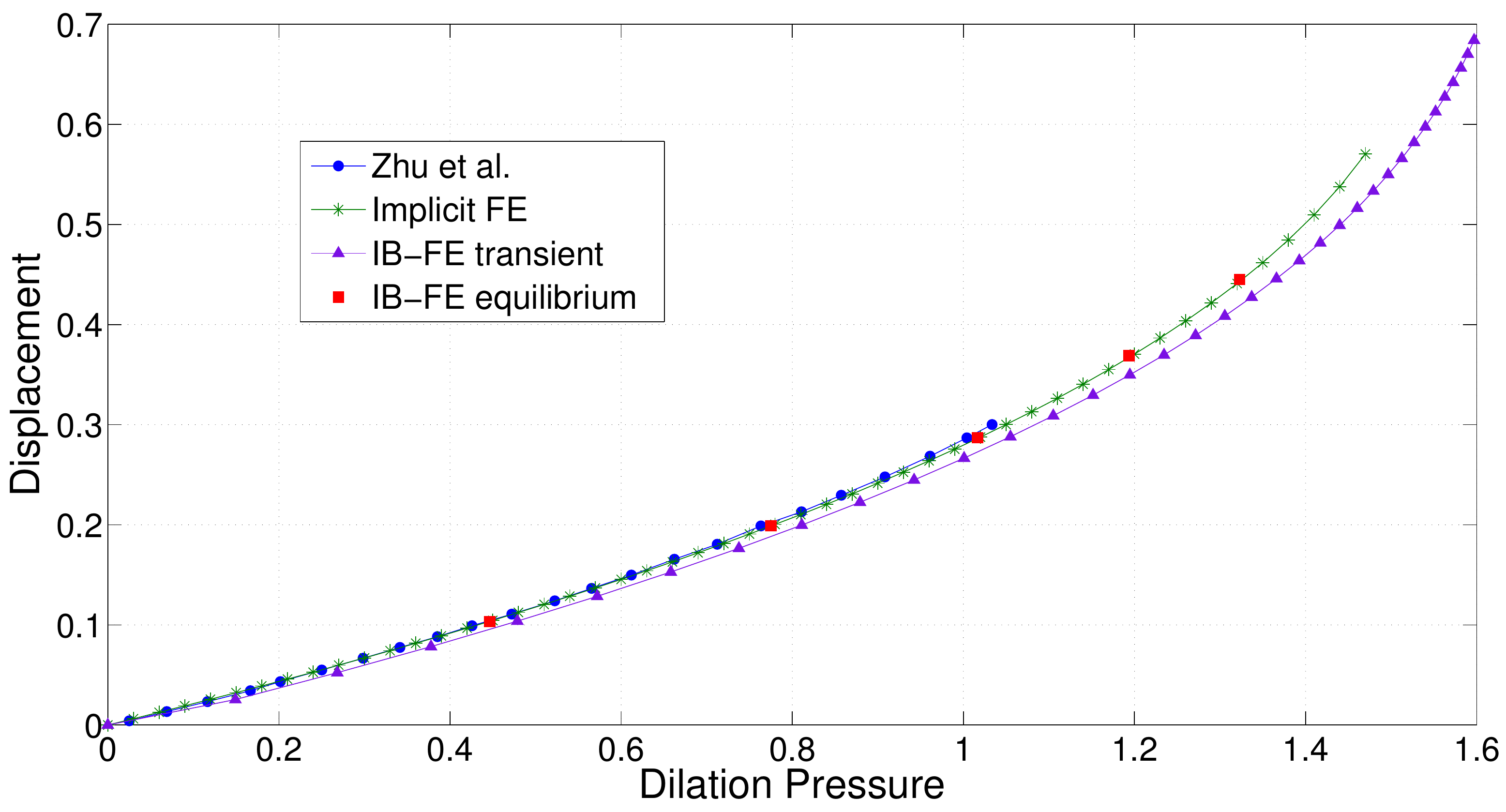} 
 \caption{Radial displacement versus dilation pressure. The radial displacement is measured at point $(R = 0.5,\Theta = 0, Z= 0.5)$. \textit{Curve Zhu et al.}: results from Zhu et al.~\cite{zhu2013three}. \textit{Curve Implicit FE}: results based on the implicit FE method. \textit{Curve IB-FE transient}: results based on our immersed boundary finite element (IB-FE) method, in which we have a non-zero fluid source during the entire simulation. \textit{Points IB-FE equilibrium}: results when the entire fluid-structure system at different dilation level achieves equilibrium. The equilibrium at a certain dilation level is achieved by first dilating the tube for some time and then turning off the fluid source to let the velocity field vanish.}
 \label{fig_validation_1_pdx}
 \end{figure}

Zhu et al.~\cite{zhu2013three} have studied a similar case, where the tube of the same geometry and elastic model is dilated under a specified inner pressure. They calculated the equilibrium based on an implicit FE method. We also developed a similar implicit FE solver (implicit FEM) based on a finite element library \textit{libMesh}~\cite{kirk2006libmesh}, in order to dilate the tube under an even larger inner pressure. To compare solutions from the IB-FE method with those from the implicit FE method, we consider two cases, referred to as the transient case and the equilibrium case, respectively. For the transient case, we run with the fluid source specified as in eq. \eqref{eqn_validation_case1_q} until the tube becomes unstable. For the equilibrium case, we first create several restarting points at different time when we run the transient case. Then at each restarting point, we ``turn off" the fluid source and re-run the simulation until we achieve an equilibrium state (i.e. the velocity almost vanishes.). Similar to Zhu et al.~\cite{zhu2013three}, we measure the radial displacement of the material point, $(R = 0.5,\Theta = 0, Z= 0.5)$ in the initial configuration. We measure the \textit{dilation pressure} as the fluid pressure near the center of the box, i.e. $(x=0,y=0,z=0.5)$. The results are shown in Fig.~\ref{fig_validation_1_pdx}. It can be seen the transient case captures the trend. Once the system comes to the equilibrium, the number also matches well. 

\begin{figure}[ht] 
 \centering
 \includegraphics[scale = 0.3]{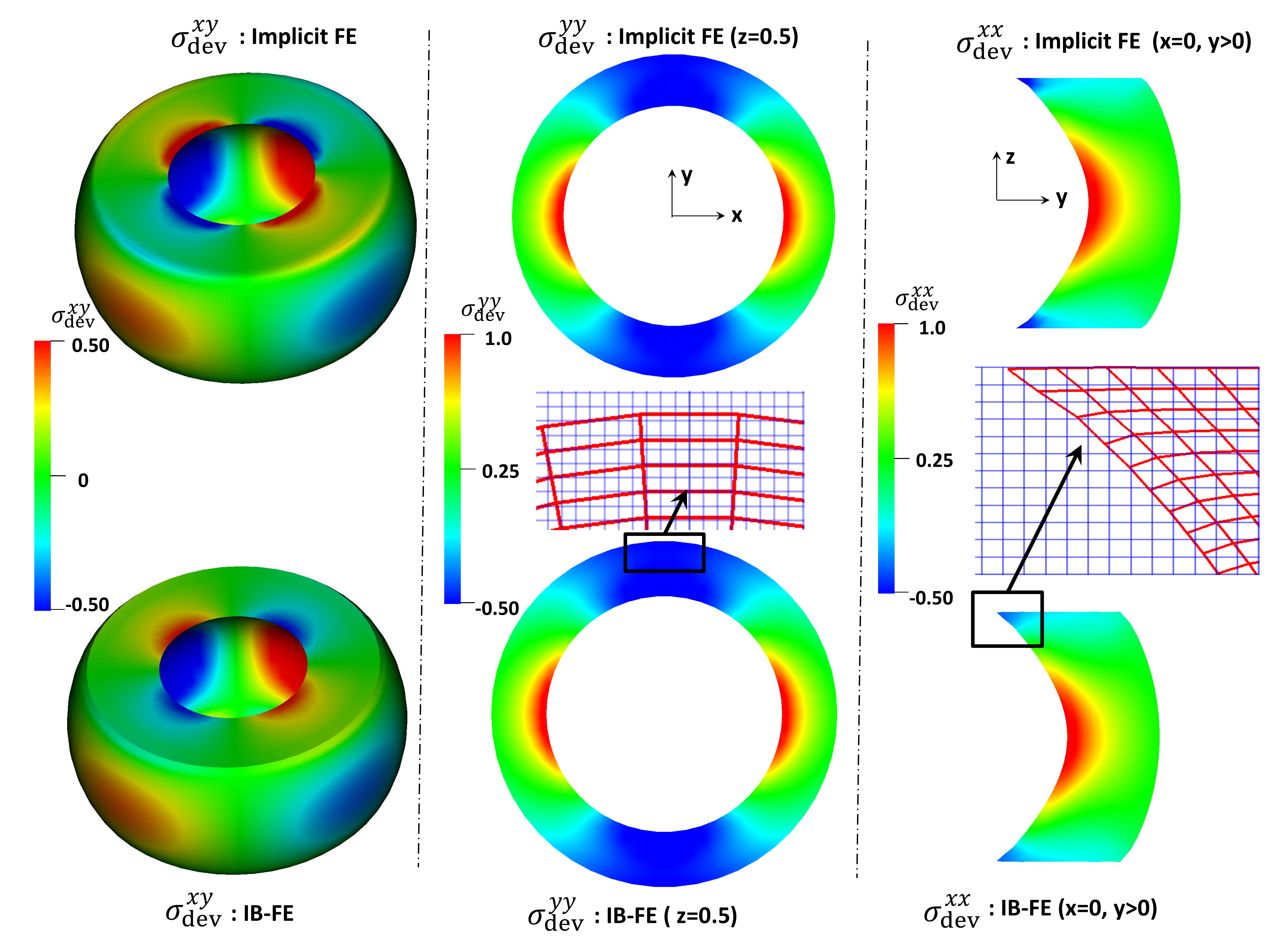} 
 \caption{Deviatoric Cauchy stress components from our IB finite element (IB-FE) method (bottom) versus those from implicit finite element (implicit FE) method (top). \textit{Left}: the predicted xy-component Deviatoric Cauchy stress. \textit{Middle}: the predicted yy-component Deviatoric Cauchy stress in plane z=0.5. The fluid mesh (light blue) and deformed solid mesh (dark red) near the top is also shown.  \textit{Right}: the predicted xx-component Deviatoric Cauchy stress in the right half plane x=0. The fluid mesh (light blue) and deformed solid mesh (dark red) near the top is also shown.}
 \label{fig_validation_1_stress}
 \end{figure}

We also compare the deviatoric Cauchy stress components obtained from our IB-FE method with those obtained from the implicit FE method, when the radial displacement at the material point, $(R = 0.5,\Theta = 0, Z= 0.5)$ for both cases is the same. This is shown in Fig.~\ref{fig_validation_1_stress}. The distribution of $\vec{\sigma}_{dev}^{xy}$, $\vec{\sigma}_{dev}^{yy}$ and $\vec{\sigma}_{dev}^{xx}$ from the two methods matches very well, except on the top and bottom surfaces of the tube. The stress difference on the two surfaces is because the IB-FE method and the implicit FE method handle the Dirichlet boundary conditions differently. The former uses the penalty method to fix the two surfaces, whereas the latter applies the Dirichlet boundary conditions directly. The fluid and solid meshes are also shown to highlight the large deformation and mesh-size ratio. We can see that the deformed configuration predicted from the two methods looks almost identical. No leakage occurs even though the circumferential mesh size of the structure near the top is around six times of the fluid mesh size. We remark that a good performance of our IB-FE method is still achieved, even though the structure is largely deformed and the Lagrangian mesh size becomes much coarser than the Eulerian mesh. 
 
\subsection{ Validation of the fiber-matrix material model: dilation of a long fiber-reinforced tube }
\label{sec_tube_dilation}
\begin{figure}[ht] 
 \centering
 \includegraphics[scale = 0.3]{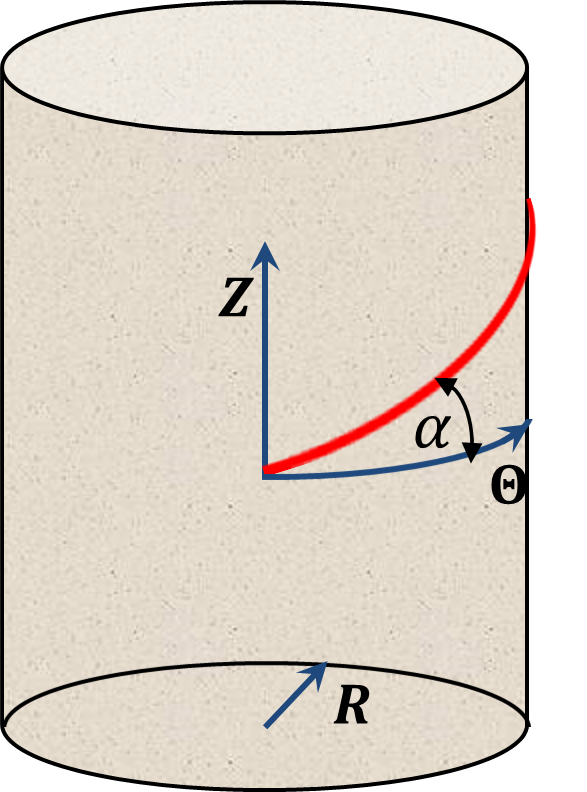} 
 \caption{Illustration of the fiber angle to characterize a fiber's orientation. A tube is described in the cylindrical coordinates $\s = (R, \Theta, Z)$. For a family of fibers running in $(\Theta, Z)$ plane, the fiber angel $\alpha$ is  measured with respect to the circumferential orientation, $\hat{\Theta}$.  }
 \label{fig_validation_2_fiber_angle}
 \end{figure}

Since the esophagus is a multiple-layered tube that consists of matrix and families of fibers, we need to examine the performance of our method in dealing with fiber-matrix material models. Hence, we present a second validation study on dilation of a long fiber-reinforced cylindrical tube. The tube is also described in the cylindrical coordinates $\s = (R, \Theta, Z)$. But the tube is much thinner and longer, and its initial stress-free configuration is $1.0 \leq R \leq 1.2, 0 \leq \Theta \leq 2\pi, 0 \leq Z \leq 10$. The tube is assumed to consist of an isotropic matrix and two families of continuous fibers running in the $\Theta, Z$ plane. We introduce the \textit{fiber angle} to characterize the orientation of a family of fibers running in $(\Theta, Z)$ plane. The fiber angle is measured with respect to the circumferential orientation, $\hat{\Theta}$, as shown in Fig.~\ref{fig_validation_2_fiber_angle}.

Specifically, if the fiber angle is $\alpha$, then the orientation of the fiber is $(0\hat{R}, \cos{\alpha} \hat{\Theta}, \sin{\alpha} \hat{Z})$.  In this study, we take the fiber angles of the two families of fibers as $\alpha_1 = \alpha$ and $\alpha_2 = 180 - \alpha$ degree, respectively. And we specify the elastic model for the matrix and fibers as below,
\begin{align}
\varPsi &= \varPsim(I_1) +\varPsi_{\fb1}(I_{\fb 1}) + \varPsi_{\fb2} (I_{\fb 2}),   \label{eqn_validation_case2_psi} \\
\varPsim(I_1) &= C_1/2  (I_1 -3),                            \label{eqn_validation_case2_psim} \\
\varPsi_{\fbi}(I_{\fbi}) &= C_2 /2 (\sqrt{I_{\fbi}} - 1) ^2, (i=1,2).  \label{eqn_validation_case2_psif}
\end{align}
Similar to the first validation case, we fix the two ends of the tube through the penalty method. The fluid domain is $-2.0 \leq x \leq 2.0, -2.0 \leq y \leq 2.0, -0.0 \leq z \leq 10.0$. The density and viscosity of the fluid is set to be 1.0. We also specify a fluid source $q(\x, t)$ to dilate the tube,
\begin{equation}
 q(\x,t) = \begin{cases} q_0 (1.0 - e^{-t}) & \mbox{if } (x^2 + y^2 > 0.4^2, z \in [0.1,9.9]), \\
 0 & \mbox{otherwise}, \end{cases} \label{eqn_validation_case2_q} 
\end{equation}
where $q_0 = 0.03125$. 
The tube is discretized using eight-node hexahedral finite elements, with the number of elements as $(n_R, n_\Theta, n_Z) = (1, 50 ,25)$. The fluid is discretized based on the finite difference method, with the mesh-size as $h_x = h_y = h_z = 0.1$. 

In this case, we compare our numerical solution with a derived analytical solution. Because fiber angles of the two fiber families sum up to 180 degree, the deformation of the tube can be shown to be axially symmetric (see Appendix B). Moreover, since the tube is very thin, the axial stretch is assumed to be uniform. Notice that this assumption is valid near the middle region of the tube. Based on those two conditions, we can derive an analytical expression for the inner pressure, $ P_{\text{inner}}$ (i.e. the pressure inside the tube), when the system achieves the equilibrium.
\begin{align}
 P_{\text{inner}} &=  \int_{R_{i}}^{R_{o}} \frac{C_1}{r} \left[\frac{r^2}{R^2} -\left (\frac{dr}{dR}\right)^2 \right ]\frac{dr}{dR} dR, \nonumber \\
 &+ \int_{R_{i}}^{R_{o}} \frac{2C_2}{r} \left(\frac{r \cos\alpha}{R} \right)^2 \left [1 - 1/\sqrt{\left (\frac{r \cos\alpha}{R}\right)^2 + (\lambda_z \sin\alpha)^2 } \right ]\frac{dr}{dR} dR. \label{eqn_validation_case2_Pinner}
\end{align} 
$r$ and $R$ are the deformed and initial radial coordinates, respectively. $R_i$ and $R_o$are the inner and outer radii, respectively. $\lambda_z$ is the axial stretch ratio. $\alpha$ is the fiber angle of one family of fibers, with the fiber angle of the other family as $180 - \alpha$. If we know the deformed inner radius and outer radius, denoted as $r_i$ and $r_o$, we can obtain the current radial coordinate $r(R)$ and axial stretch $\lambda_z$ as below,
\begin{align}
 r(R) &= \sqrt{R^2 \left ( \frac{r_o^2 -r_i^2 }{R_o^2 -R_i^2} \right ) + r_i^2 - R_i^2 \left( \frac{r_o^2 -r_i^2 }{R_o^2 -R_i^2} \right ) }, \label{eqn_validation_case2_rR} \\
 \lambda_z&=  \frac{R_o^2 -R_i^2}{r_o^2 -r_i^2 }. \label{eqn_validation_case2_lambdaz} 
\end{align}
The details on the derivation of eqs. \eqref{eqn_validation_case2_Pinner}-\eqref{eqn_validation_case2_lambdaz} are given in the Appendix B.

We simulate cases with different fiber angles $\alpha$. For each fiber angle, we first obtain a transient case with the source term~\eqref{eqn_validation_case2_q}. We then restart simulations with zero source term at different restarting points to obtain several equilibrium states. At each equilibrium state, we measure the $r_i$, $r_o$, and $\lambda_z$ in the middle of the tube, and the inner pressure, denoted as $P_\text{numerical}$. We then compute the analytical inner pressure, denoted as $P_\text{analytic}$ based on measured parameters and eqs. \eqref{eqn_validation_case2_Pinner}-\eqref{eqn_validation_case2_lambdaz}. We compare $P_\text{numerical}$ and $P_\text{analytic}$, as listed in Table \ref{table_validation_case2_Pinner}. It can be seen, the errors in most cases are below 0.1 percent and in worst cases are below 4 percent.

\begin{table}[h]
 \caption{Error in the inner pressure for cases with different fiber angles, at different dilation levels. $r_i$ and $r_o$ is the deformed inner and outer radius in the middle section of the tube, respectively. $P_\text{numerical}$ is the inner pressure in the middle section of the tube obtained from our simulations. $P_\text{analytic}$ is the inner pressure based on eq. \eqref {eqn_validation_case2_Pinner}. The relative error $\epsilon_p = \frac{|P_{\text{numerical}} - P_\text{analytic}|}{P_\text{analytic}}$}
 \centering
\begin{tabular}{l | l | l | l | l | l | l}
 \hline \hline
 Fiber angle $\alpha$ (degree)& $r_i$  & $r_o$ & $P_{\text{analytic}}$  &   $P_\text{numerical}$ & $\epsilon_p$\\ [1ex]
 \hline
 45  			      & 1.1345 & 1.313 & 0.0857 	        & 0.0830 		& 3.15e-2\\
			      & 1.2636 & 1.428 & 0.1387 		& 0.1388 		& 7.21e-4\\
			      & 1.3850 & 1.537 & 0.1781 		& 0.1777 		& 2.25e-3\\
 \hline
 60  			      & 1.1332 & 1.312 & 0.0701 	        & 0.0676 		& 3.57e-2\\
			      & 1.2632 & 1.427 & 0.1111 		& 0.1105 		& 5.40e-3\\
			      & 1.3875 & 1.539 & 0.1390 		& 0.1388 		& 1.44e-3\\
 \hline
 0			      & 1.1312 & 1.310 & 0.1355 		& 0.1343 		& 8.86e-3\\
			      & 1.2524 & 1.417 & 0.2271			& 0.2264		& 3.08e-3\\
 \hline
 \end{tabular}
 \label{table_validation_case2_Pinner}
\end{table}

\section{Esophageal transport}
\label{part4_model}
In Section~\ref{part3_validation}, we showed that our IB-FE method is able to accurately handle the fluid-structure interaction that involves very large deformation, and fiber-matrix interactions. It performs well with no leakage even though the Lagrangian mesh becomes much coarser than the fluid mesh. In this section, we proceed with our main application, which is a three-dimensional continuum-based model on esophageal transport.

Esophageal transport is a bio-physical process that transfers a food bolus from the pharynx to the stomach through the esophageal tube. It involves the interaction among food bolus, multi-layered esophageal wall and neurally-controlled muscle activation. The bolus is generally treated as a Newtonian fluid, with its viscosity varying from one centipoise (cP) to several hundred centipoise~\cite{Dooley1988}. The multi-layered esophageal wall consists of inner mucosal-submucosal layer (collectively referred to as ``mucosal" layer), circular muscle layer, and longitudinal muscle layer (so named because of their fiber orientations~\cite{Kahrilas1989}). The neurally-controlled muscle activation involves the contraction of muscle fibers in both circular muscle (CM) layer and longitudinal muscle (LM) layer, which is referred to as \textit{circular muscle (CM) contraction} and \textit{longitudinal muscle (LM) shortening}, respectively. The bio-physical process we attempt to model is: how does a bolus transport from the esophageal top to the esophageal bottom under the neurally-controlled muscle activation? This is the same process simulated in our previous work~\cite{kou2015fully}, but here we employ a continuum-based material model, in order to consider more realistic and complicated material behavior of the esophageal tissue. 

\subsection{Geometry, boundary conditions and material properties}

The esophagus in the reference configuration is taken to be a long straight cylindrical tube. The geometry of the esophageal tube in this model is the same as our previous model~\cite{kou2015fully}, except that the esophagus's length is taken to be 180 mm. We reduce the esophagus's length from 240 mm to 180 mm in this model, because the esophageal tube modeled here does not include the lower and upper sphincters and thus is shorter than the entire esophagus. We model the esophageal wall as a three-layered composite, including the mucosa, CM and LM layers. We consider a thin liquid layer lining along the esophageal lumen when the esophagus is at rest, similar to previous models~\cite{kou2015fully,Li-Brasseur1993,Ghosh2005}. We calculate the thickness of each esophageal layer based on clinical data~\cite{Mittal2005}. The lumen radius at rest is 0.3 mm. The thickness of mucosal, CM, and LM layers are 3.8 mm, 0.6 mm, and 0.6 mm, respectively. 

\begin{figure}[ht] 
 \centering
 \includegraphics[scale = 0.28]{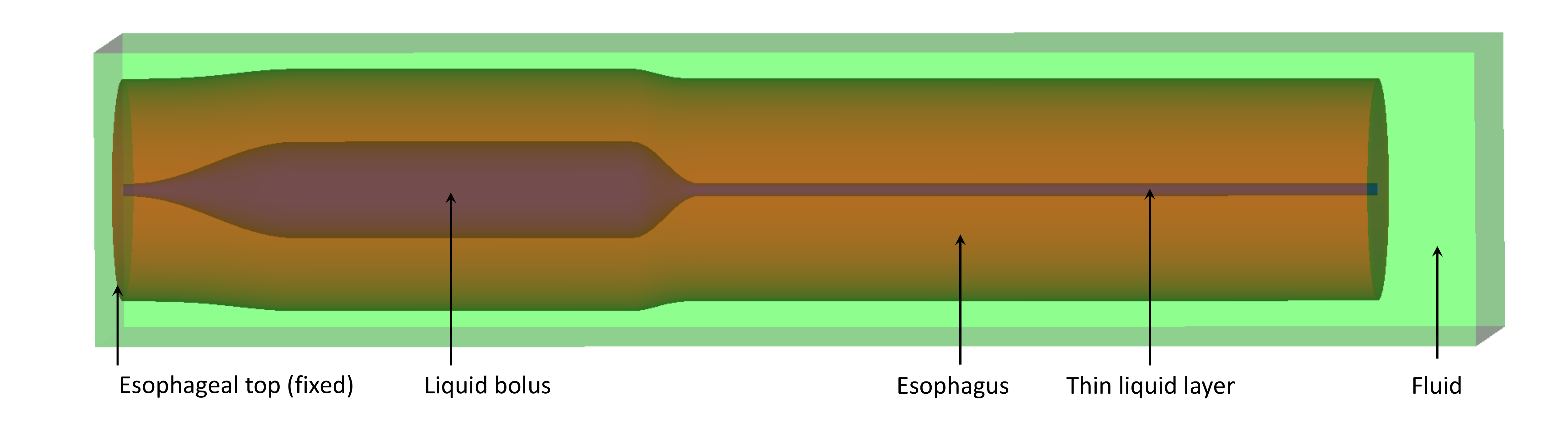} 
 \caption{Schematic of the computational domain for the esophageal transport model. The elastic esophagus, a cylindrical tube is immersed in a background fluid box. The esophageal top is fixed. The upper esophagus is initially filled with a bolus and the lower part is filled with a thin liquid layer in the lumen. The top surface of the rectangular computational domain has zero-velocity boundary conditions. All the other five surfaces have traction-free boundary conditions.  }
 \label{fig_overall_setup}
 \end{figure}

To use the IB-FE method, we immerse the entire esophagus in a fluid box of size (-7 mm, 7 mm) x (-7 mm, 7 mm) x (-126 mm, 180 mm). On the top surface of the fluid box, we specify zero-velocity boundary conditions to fix the esophageal top. This corresponds to the physiological constrains from the upper esophageal sphincter, which fixes the esophageal top in place. We also specify a penalty term to better fix the esophageal top. On the other surfaces of the fluid box, we impose traction-free boundary conditions. We simulate the transport of a bolus filled in the upper esophageal body initially. A schematic of the overall setup is shown in Fig.~\ref{fig_overall_setup}.

As for the material property, we consider fluid and all the esophageal layers to assume the same viscosity of 10 cP and the same density of 1 g/cm$^3$. We also need to include the elastic property of each esophageal layer based on experiments. In-vitro experiments on material properties show that the esophageal tissue can be generally characterized as a nonlinear anisotropic elastic or pseudo-elastic material. However, quite different material models have been proposed among different groups~\cite{Yang2006a,Yang2006b,Natali2009,Sokolis2013,Stavropoulou2009}. Yang et al.~\cite{Yang2006a} have proposed a so-called bi-linear model to characterize both the mucosal and muscle layers. Each esophageal layer is modeled as a fiber-reinforced material that consists of ground tissue (matrix) and elastic fibers. Natali et al.~\cite{Natali2009} also adopts a fiber-reinforced model to characterize mucosal and muscle layers, but the model is of an exponential form. Stavropoulou et al.~\cite{Stavropoulou2009} adopted Fung-type material models to characterize the mucosa and muscle layers, in which fiber model is not included. On the other hand, experiments on the histological information suggest both the mucosa and muscle layers are biological tissues embedded with biological fibers \cite{Natali2009}. Therefore, we prefer to model esophageal layers as fiber-reinforced materials, especially when we need to include the active contraction/relaxation of muscle fibers. To test the capabilities of our model, we present three cases here. The first case is based on a bi-linear model proposed in~\cite{Yang2006a}. The second case is based on an exponential model proposed in~\cite{Natali2009}. The third case is also based on a bi-linear model, but it includes more realistic and complicated muscle fiber architecture. The bi-linear model and the exponential model are discussed as below. Notice that the two models in their original forms include only two layers: mucosal and muscle layers. In our model, the muscle layers is split into two layers: CM and LM layers, to be consistent with more detailed histological information. For the material model on muscle fibers, we also introduce one additional parameter, referred here as the \textit{reference stretch ratio} of fibers. This parameter is used to model neurally-controlled muscle activation, which will be discussed later.

\subsubsection{Bi-linear model}
\label{mod1_bilinear}
This model is based on Yang et al.~\cite{Yang2006a}, in which they refer to the model as the bi-linear model. First is the mucosal layer. We remark that we adopt a different material model on mucosal layer. This is because the material model proposed in Yang et al.~\cite{Yang2006a} is based on an in-vitro test. However, the in-vivo mucosal layer is substantially different from the in-vitro one in both the geometry and material behavior. The in-vivo mucosal layer is highly folded with a residual stress~\cite{Greg2000}. Its stiffness along the lateral direction is very low so that the esophageal tube attains a high distensibility, as seen in the clinical endoscopy, whereas the axial stiffness of mucosal layer is relatively high. Thus we model the mucosal layer here as a composite that is reinforced by a family fiber along the axial direction. The material model is as below, 
\begin{align}
\varPsi^{\tmo}& = \varPsim^{\tmo} + \varPsif^{\tmo}, \\
\varPsim^{\tmo} &= \frac{C_0}{2}(I_1 -3), \\
\varPsif^{\tmo} &=\frac{C_1}{2} \left[\left(\sqrt{I_{\fb}^{\tmo}} -1 \right ) ^2 \right ],
\end{align}
where $I_{\fb}^{\tmo} = \tensor{C} : (\vec{a}^{\tmo} \bigotimes \vec{a}^{\tmo})$. $\vec{a}^{\tmo} = (0\hat{R}, 0 \hat{\Theta}, 1 \hat{Z})$, as the fiber angle of the axial fibers in the mucosal layer is 90 degree (i.e. along the axial direction).

Second is the CM layer. Its material model is as below,
\begin{align}
\varPsi^{\tcm}& = \varPsim^{\tcm} + \varPsif^{\tcm}, \\
\varPsim^{\tcm} &= \frac{C_2}{2}(I_1 -3), \\
\varPsif^{\tcm} &= \frac{C_3}{2} \left [ \left ( \frac{ \sqrt{I_{\fb}^{\tcm}} }{ \lambda^{\tcm}} -1 \right ) ^2 \right ],
\end{align}
where $I_{\fb}^{\tcm} = \tensor{C} : (\vec{a}^{\tcm} \bigotimes \vec{a}^{\tcm})$. $\vec{a}^{\tcm} = (0\hat{R}, \cos{\alpha^{\tcm}} \hat{\Theta}, \sin{\alpha^{\tcm}} \hat{Z})$. $ \alpha^{\tcm} $ is the fiber angle of the circular muscle fibers. $\lambda^{\tcm}$ is the reference stretch ratio that is included to deal with circular muscle fiber contraction.

Third is the LM layer. Its material model is as below,
\begin{align}
\varPsi^{\tlm}& = \varPsim^{\tlm} + \varPsif^{\tlm}, \\
\varPsim^{\tlm} &= \frac{C_5}{2}(I_1 -3), \\
\varPsif^{\tlm} &= \frac{C_6}{2}\left [ \left ( \frac{\sqrt{I_{\fb}^{\tlm}} }{ \lambda^{\tlm}} -1 \right ) ^2 \right ],
\end{align}
where $I_{\fb}^{\tlm} = \tensor{C} : (\vec{a}^{\tlm} \bigotimes \vec{a}^{\tlm})$. $\vec{a}^{\tlm} = (0\hat{R}, \cos{\alpha^{\tlm}} \hat{\Theta}, \sin{\alpha^{\tlm}} \hat{Z})$. $ \alpha^{\tlm} $ is the fiber angle of the circular muscle fibers.

\subsubsection{Exponential model}
\label{mod2_exp}
Here we use a fiber-reinforced material model from Natali et al.~\cite{Natali2009} and we refer to this model as the exponential model. 
First is the mucosal layer for which we include a family of axial fibers to consider high stiffness along the axial direction. Notice that since the original model has two families of fibers in the mucosal layer, we now have three families of fibers in our model on the mucosal layer.
\begin{align}
\varPsi^{\tmo}& = \varPsim^{\tmo} + \varPsif^{\tmo}, \\
\varPsim^{\tmo} &=  \frac{C_0}{2}(I_1 -3),\\
\varPsif^{\tmo} &= \frac{C_1}{2} \left [ \left (\sqrt{I_{\fb 1}^{\tmo} } -1 \right ) ^2 \right ] \nonumber \\
		& + \frac{C_2 }{k_2 ^2}  \left [ e ^ {k_2 \left (\frac{I_{\fb 2}^{\tmo}}{(\lambda_2^{\tmo})^2} -1 \right ) } -k_2 (I_{\fb 2}^{\tmo} -1) - 1  \right ]  \nonumber \\
		& + \frac{C_3 }{k_3 ^2}  \left [ e ^ {k_3 \left (\frac{I_{\fb 3}^{\tmo}}{(\lambda_3^{\tmo})^2} -1 \right ) } -k_3 (I_{\fb 3}^{\tmo} -1) - 1  \right ],  
\end{align}
where $I_{\fbi}^{\tmo} = \tensor{C} : (\vec{a}_i^{\tmo} \bigotimes \vec{a}_i^{\tmo})$. $\vec{a}_i^{\tmo} = (0\hat{R}, \cos{\alpha_i^{\tmo}} \hat{\Theta}, \sin{\alpha_i^{\tmo}} \hat{Z})$. $ \alpha_i^{\tmo} $ is the fiber angle. In particular, $ \alpha_1^{\tmo} = 90$ degree, same as the bi-linear model. $\lambda_i^{\tmo}$ is the reference stretch ratio under which the fiber elastic potential is zero.

Second is the CM layer. The model includes ground tissue and one family of the circular muscle fibers as below,
\begin{align}
\varPsi^{\tcm}& = \varPsim^{\tcm} + \varPsif^{\tcm}, \\
\varPsim^{\tcm} &= \frac{C_4}{ k_4} \left (e ^ {k_4 (I_1 -3)} -1 \right), \\
\varPsif^{\tcm} &= \frac{C_5 }{k_5 ^2}  \left [ e ^ {k_5 \left (\frac{I_{\fb}^{\tcm}}{(\lambda^{\tcm})^2} -1 \right ) } -k_5 (I_{\fb}^{\tcm} -1) - 1  \right ],  
\end{align}
where $I_{\fb}^{\tcm} = \tensor{C} : (\vec{a}^{\tcm} \bigotimes \vec{a}^{\tcm})$. $\vec{a}^{\tcm} = (0\hat{R}, \cos{\alpha^{\tcm}} \hat{\Theta}, \sin{\alpha^{\tcm}} \hat{Z})$. $ \alpha^{\tcm} $ is the fiber angle of the circular muscle fibers. $\lambda^{\tcm}$ is the reference stretch ratio that is included to deal with circular muscle fiber contraction.

Third is the LM layer. The model includes ground tissue and one family of the longitudinal muscle fibers as below,
\begin{align}
\varPsi^{\tlm}& = \varPsim^{\tlm} + \varPsif^{\tlm}, \\
\varPsim^{\tlm} &= \frac{C_6}{ k_6} \left (e ^ {k_6 (I_1 -3)} -1 \right), \\
\varPsif^{\tlm} &= \frac{C_7 }{k_7 ^2}  \left [ e ^ {k_7 \left (\frac{I_{\fb}^{\tlm}}{(\lambda^{\tlm})^2} -1 \right ) } -k_7 (I_{\fb}^{\tlm} -1) - 1  \right ],  
\end{align}
where $I_{\fb}^{\tlm} = \tensor{C} : (\vec{a}^{\tlm} \bigotimes \vec{a}^{\tlm})$. $\vec{a}^{\tlm} = (0\hat{R}, \cos{\alpha^{\tlm}} \hat{\Theta}, \sin{\alpha^{\tlm}} \hat{Z})$. $ \alpha^{\tlm} $ is the fiber angle of the longitudinal muscle fibers. $\lambda^{\tlm}$ is the reference stretch ratio that is included to deal with longitudinal muscle fiber contraction.

\subsection{Muscle activation}
\label{sec_muscle_active}
Neurally-controlled muscle activation provides the pumping force for bolus transport when the gravitational assistance is minimal. Two types of muscle activations, CM contraction and LM shortening, are involved as observed in studies based on in-vivo experiments~\cite{Kahrilas1997,Mittal2006}. CM contraction and LM shortening occur as two synchronized traveling waves in the normal physiology. The two waves originate from the sequential contraction and relaxation of corresponding muscle fibers. However, the underlying process of neuronal firing or reaction kinematics at the continuum scale is still not available. Our previous fiber-based method proposed a muscle-activation model by dynamically changing the rest lengths of springs that represent muscle fibers. Inspired by the success, here we apply the same idea to the continuum-based method. We model the muscle activation by dynamically changing the reference stretch ratio of corresponding muscle fibers. Specifically, let $Z$ denote the vertical coordinate in the reference configuration of the esophageal tube. The bottom end of the esophagus is at the origin $Z=0$, and the top is at $Z=L$. The reference stretch ratio of a muscle fiber, denoted as $(\lambda^{\text{muscle}}(Z,t),\text{muscle = CM or LM})$ is given by
\begin{equation}
\lambda^{\text{muscle}}(Z,t) = \begin{cases} 1 & \mbox{if } t-t_0 \leq \frac{L-Z}{c}, \\ 
1-a^{\text{muscle}}(Z,t) & \mbox{if } \frac{L-Z}{c} < t-t_0 < \frac{L-Z}{c} + \frac{\Delta L}{c}, \\
1 & \mbox{if }  t- t_0 \geq \frac{L-Z}{c} + \frac{\Delta L}{c}, \end{cases} \label{eqn_contract_fiber} 
\end{equation}
where $c$ is the speed of the activation wave, $t_0$ is the initiation time of activation, and $\Delta L$ is the contracting segment's length in the reference configuration. Eq.~\eqref{eqn_contract_fiber} gives the reference stretch ratio of a fiber at its rest, activation, and relaxation states, respectively. The equation also shows that at any time, the whole esophageal tube has a contracting segment with a vertical length $\Delta L$. $a^{\text{muscle}}(Z,t)$ is referred to as the reduction ratio, whose form is the same as our previous model~\cite{kou2015fully},
\begin{equation}
 a^{\text{muscle}}(Z,t)= a_0^{\text{muscle}} e^{-0.5(Z-Z_0(t))^2/{W}^2}, \label{eqn_nonuniform_contraction_fiber}
\end{equation}
where $a_0^{\text{muscle}}$ is a constant reduction ratio, $Z_0(t) = c (t - t_0)$ is the $Z$-coordinate of the vertical center of the contraction segment, and $W$ is the parameter that controls the width of the Gaussian distribution in eq. \eqref{eqn_nonuniform_contraction_fiber}.

The common parameters of muscle activation model used in the two cases of esophageal transport are listed in Table~\ref{tab_activation_para_fiber}.
\begin{table}[ht]
 \caption{Model parameters for the circular muscle (CM) contraction and longitudinal muscle (LM) shortening used in all the cases. The muscle 
 activation model is based on eqs.~\eqref{eqn_contract_fiber} and ~\eqref{eqn_nonuniform_contraction_fiber} . }
 \centering
\begin{tabular}{l| l | l | l | l | l | l }
 \hline \hline
 Muscle activation type  & $a_0$& $c$ (mm/s)  & $\Delta L$ (mm)& $W$ (mm) & $t_0$ (s)  \\ [1ex]
 \hline
 CM contraction  &0.4 & 100 & 60 & 15 & 0\\
 LM shortening  &0.4 & 100 & 60 & 15 & 0\\ 
 \hline
 \end{tabular}
 \label{tab_activation_para_fiber}
\end{table}

\subsection{Numerical issues}
\label{sec_num_parameter}
Esophageal transport model includes multiple length scales. This is evidenced by the fact that 
the esophageal length is 180 mm, while the lumen radius at rest is only 0.3 mm. The requirement of resolving 
the narrow lumen dictates the mesh size of the fluid grid. Moreover, very large deformation of a typical esophageal segment occurs when the bolus first comes in and then leaves. To deal with the leakage issue associated with large deformation of the Lagrangian mesh, our previous work employed a much refined mesh for inner-most esophageal layers. In this work, instead, we adopt the adaptive interaction quadrature rule that we discussed in Section~\ref{part2_formulation}. This approach permits us to use a relatively simple and coarse Lagrangian mesh for the solid. The Lagrangian mesh is based on the cylindrical coordinate system, with the mesh information of each layer listed in Table ~\ref{tab_solid_mesh_size}. The fluid mesh is a Cartesian mesh, with $h_x = h_y=0.2 ~\text{mm and } h_z =0.9 ~\text{mm}$. The time step $\Delta t$ needs to satisfy the stability constraints from both the fluid and solid systems. Based on empirical tests, we choose $\Delta t = 0.02$~ms for cases with the bi-linear model, and $\Delta t = 0.01$~ms for cases with the exponential model. The total physical time for the simulation is about 2.4 s. The code is compiled based on IBAMR: An adaptive and distributed-memory parallel implementation of the immersed boundary method~\cite{GriffithIBAMR}. All cases run on the Northwestern super-computer, Quest, with 48 processors. Cases with the bi-linear model and the exponential material model take around 160 hours and 300 hours to finish, respectively. We also conduct speed tests on cases with the present anisotropic adaptive interaction quadrature rule and the previous isotropic adaptive interaction quadrature rule using various number density. The results are listed in Table~\ref{tab_quadrature_rule}. It can be seen that cases with the anisotropic adaptive interaction quadrature rule run much faster than cases with the isotropic adaptive interaction quadrature rule, as the former cases compute much fewer interaction quadrature points per time step. In the current work, we use number density as 1.0. Using this number density, the case with the anisotropic adaptive interaction quadrature rule runs about twice faster than the case with the isotropic adaptive interaction quadrature rule.
\begin{table}[ht]
 \caption{Grid number along $(R, \Theta, Z)$ orientations, denoted as $(n_R, n_\Theta, n_Z)$, for each layer of the esophagus in the reference configuration.}
 \centering
\begin{tabular}{l | l | l | l }
 \hline \hline
 Grid number & Mucosal Layer & CM & LM \\ [1ex]
 \hline
  $n_R$      & 6 	     & 1  & 1 \\
 $n_\Theta$  & 16   	     & 16 & 16 \\ 
 $n_Z$ 	     & 180           &180 &180\\ 
 \hline
 \end{tabular}
 \label{tab_solid_mesh_size}
 
\end{table}
\begin{table}[ht]
 \caption{Speed tests for cases with different interaction quadrature rules and the different number density based on the esophageal transport model. All cases run on the Northwestern super-computer, Quest, with 48 processors. Note that both the number of the interaction quadrature points spread per time step and number of time steps advanced per hour vary during the simulation. The reported values are approximations to the average values. }
 \centering
\begin{tabular}{p{3cm} |p{2cm} |p{4cm} | p{2cm} }
 \hline \hline
 Type of adaptive interaction quadrature rule & number density & NO. of interaction quadrature points spread per time step (million) & NO. of time steps advanced per hour \\ [1ex]
 \hline
  Anisotropic rule      & 1.0	     & 2.8  	& 750 \\
   						& 1.5    & 10 		& 400 \\ 
 	     				& 2.0      & 22 		& 225 \\ 
 	     				
 \hline
 Isotropic rule			& 1.0		& 22		&320\\
 						& 1.5	& 54		&250 \\
 						& 2.0		& 108		&130 \\
 \hline
 						
 \end{tabular}
 \label{tab_quadrature_rule}
\end{table}

\subsection{Results}

\subsubsection{Case 1: Esophageal transport using the bi-linear material model}
\label{sec1_bilinear_axial}
\begin{figure}[ht] 
 \centering
 \includegraphics[scale = 0.5]{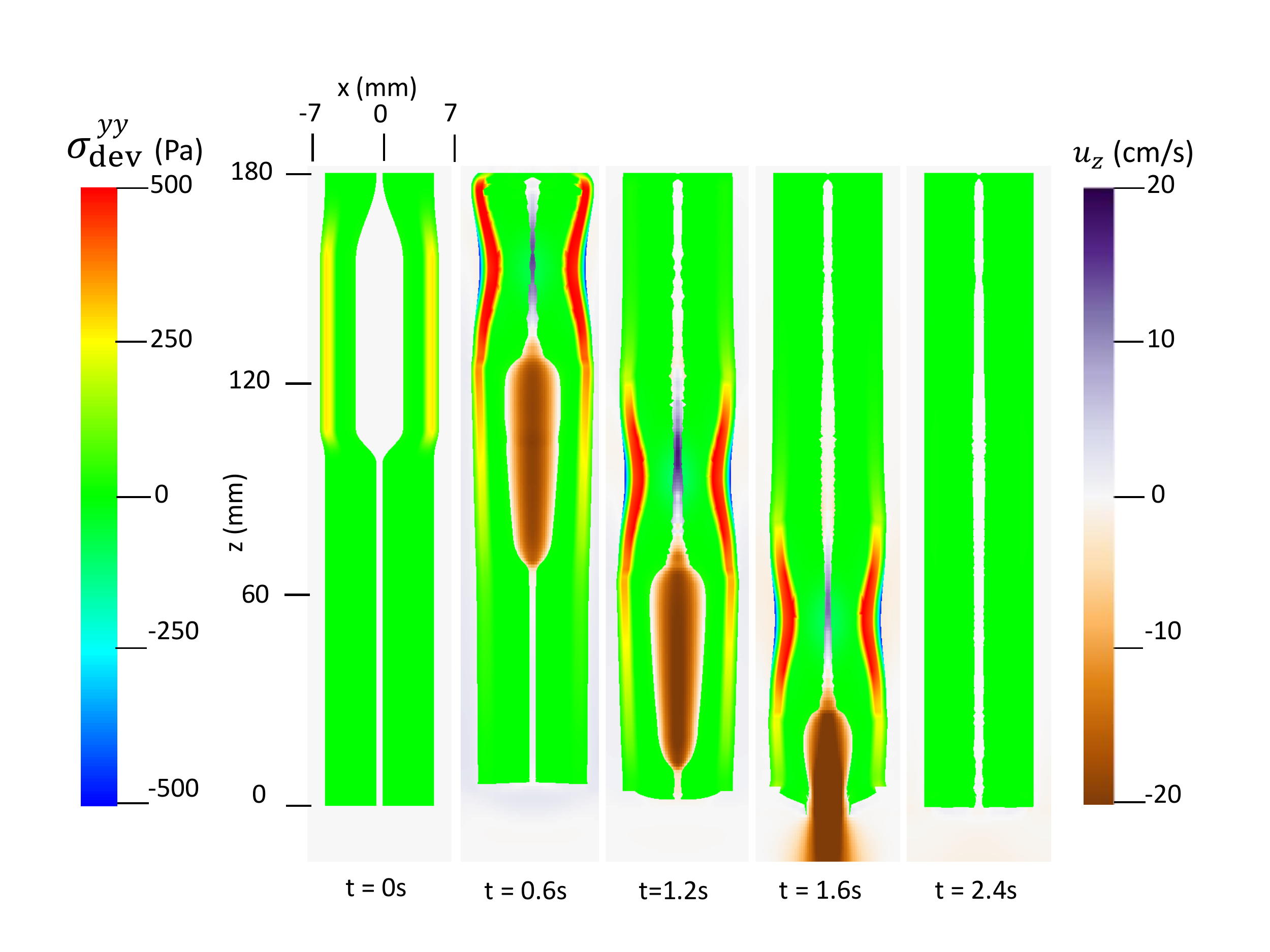} 
 \caption{Axial velocity of the bolus, $u_z$, and the yy-compoent of the deviatoric stress of the esophageal wall, $\vec{\sigma}_{dev}^{yy}$, in the plane y = 0 at different times for Case 1 in Section~\ref{sec1_bilinear_axial}. }
 \label{fig_case1_sigma_uz}
 \end{figure}

\begin{figure}[ht] 
 \centering
 \includegraphics[scale = 0.5]{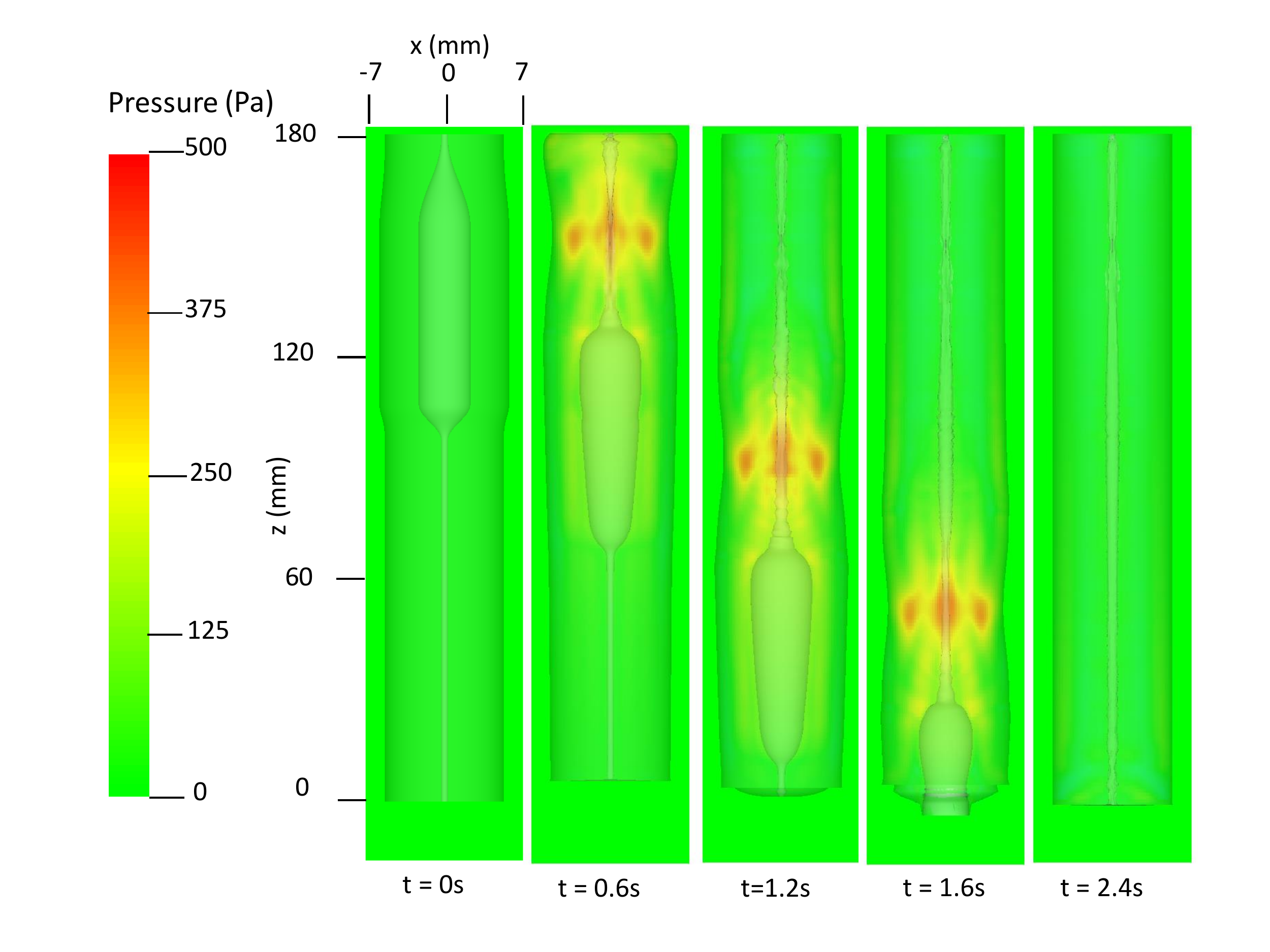} 
 \caption{Pressure field in the plane y = 0 at different times for Case 1 in Section~\ref{sec1_bilinear_axial}.}
 \label{fig_case1_pressure}
 \end{figure}

\begin{figure}[ht] 
 \centering
 \includegraphics[scale = 0.28]{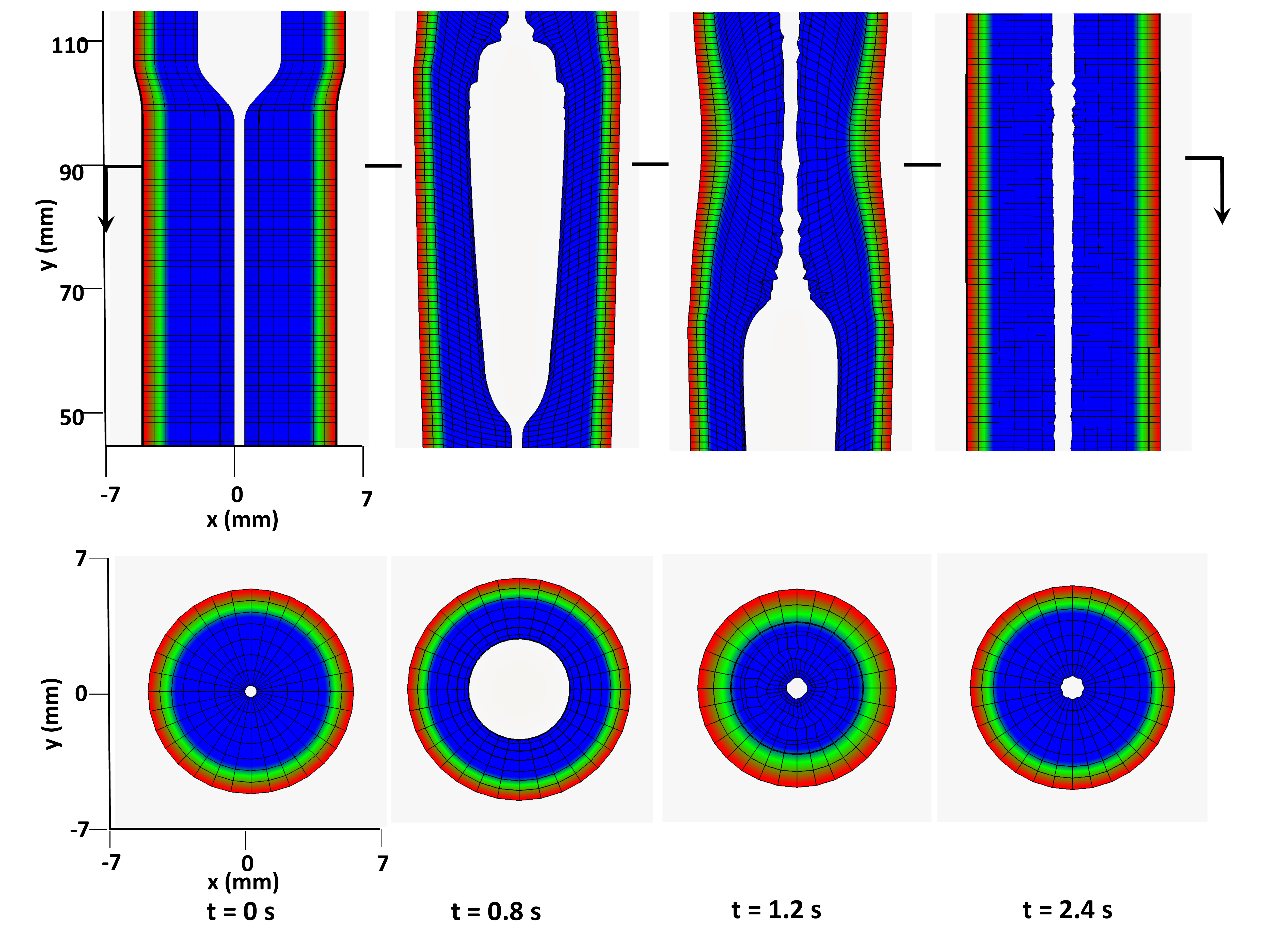} 
 \caption{Kinematics of the esophageal layers at four different stages: at rest (t = 0 s); at dilation (t = 0.8 s); at contraction (t = 1.2 s); and at relaxation (t = 2.4 s) for Case 1 in Section~\ref{sec1_bilinear_axial}. The three layers included in the model, from the inside to the outside, are the mucosa, CM, and LM layers, respectively. (Upper) Side view of a section of the esophagus within the box: (-7 mm, 7 mm) x (-0.2 mm, 0.2 mm) x (45 mm, 115 mm); (Lower) top view of a section of the esophagus within the box: (-7 mm, 7 mm) x (-7 mm, 7 mm) x (89.5 mm, 90.5 mm).  }
 \label{fig_case1_kinematics}
 \end{figure}

\begin{figure}[ht] 
 \centering
 \includegraphics[scale = 0.4]{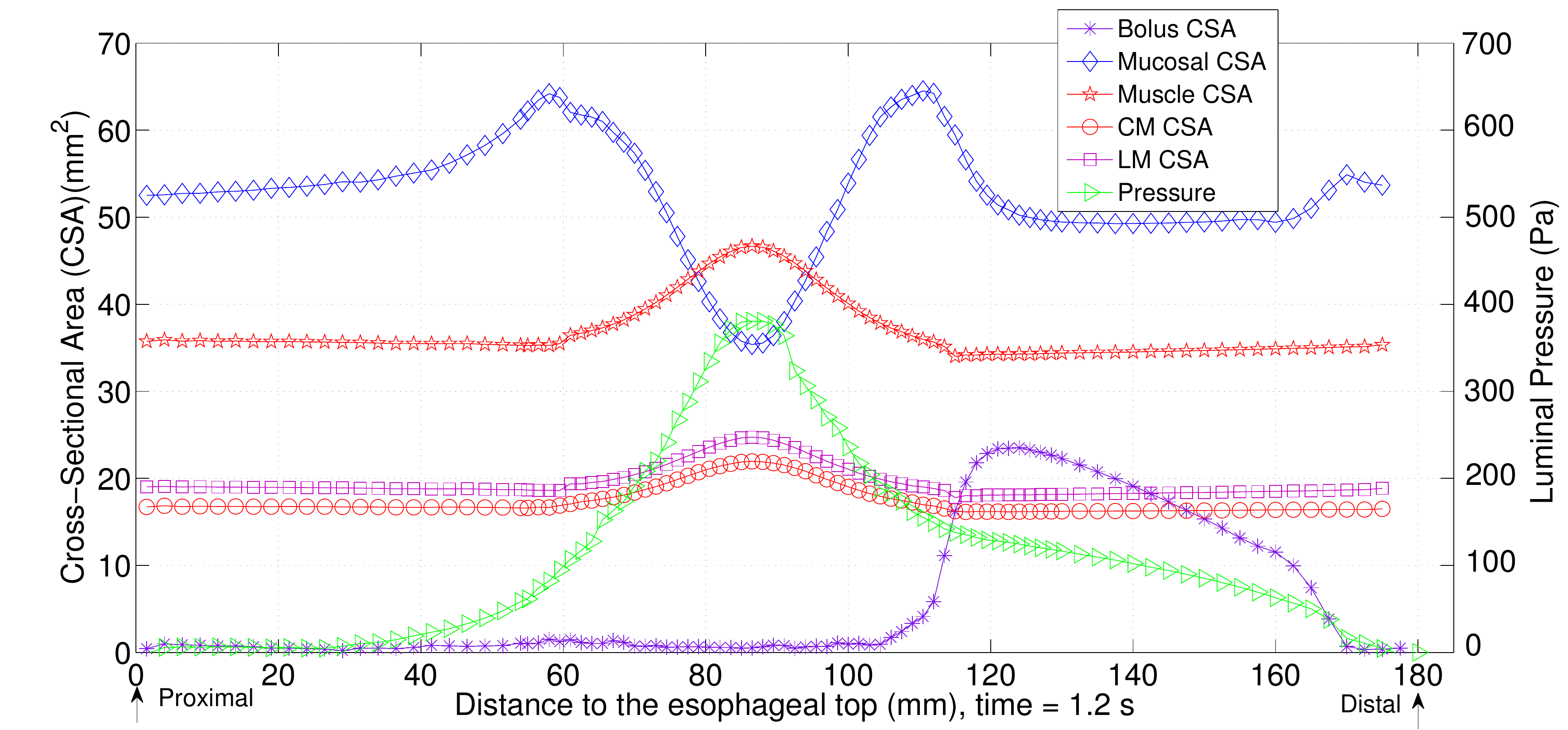} 
 \caption{The cross-sectional area (CSA) of the bolus and the esophageal components, and the lumen pressure along its central line: x = 0, y = 0, at t = 1.2 s for Case 1 in Section~\ref{sec1_bilinear_axial}.}
 \label{fig_case1_csa_p}
 \end{figure}

\begin{table}[ht]
 \caption{Model parameters of the bi-linear model (i.e. Section~\ref{mod1_bilinear}).}
 \centering
\begin{tabular}{l | l  l | l  l }
 \hline \hline
 Material type & Material parameters  \\ [1ex]
 \hline
 Mucosa & $C_0$(KPa)& 4.0e-3 & $C_1$(KPa) & 4.0e-2 \\
 \hline
 CM & $C_2$(KPa)& 4.0e-1 & $C_3$(KPa) & 4.0 \\ 
 	& $\alpha^{\tcm}$(Deg.) &0 \\
 \hline
 LM & $C_5$(KPa)& 4.0e-1 & $C_6$(KPa) & 4.0 \\ 
 	& $\alpha^{\tlm}$(Deg.) &90 \\
 \hline
 \end{tabular}
 \label{tab_bilinear_model_parameters}
\end{table}
The first case adopts the bi-linear model (i.e. Section~\ref{mod1_bilinear}) to describe the material property of esophageal tissues. Parameters for the material property are listed in Table~\ref{tab_bilinear_model_parameters}. Those parameters are adopted based on our previous fiber-based model. The originally reported parameters from in-vitro experiments are not used, as those parameters yield much stiffer muscle and mucosal layers than those in the physiological conditions~\cite{Ghosh2005,Ghosh2008}. This is likely because the in-vitro material property is substantially different from the in-vivo one. Simulation results of this case are shown in Figs.~\ref{fig_case1_sigma_uz}-\ref{fig_case1_csa_p}. They are explained as follows. Fig.~\ref{fig_case1_sigma_uz} shows both the axial velocity and the yy-component of the deviatoric stress of the esophageal wall, $\vec{\sigma}_{dev}^{yy}$. Note that, $\vec{\sigma}_{dev}^{yy}$ in the plane y=0, is also the circumferential deviatoric stress. Thus $\vec{\sigma}_{dev}^{yy}$ in CM layer characterizes the force from the CM contraction. Fig.~\ref{fig_case1_sigma_uz} clearly shows that a traveling wave of muscle activation drives a bolus downward. At about t=2.4 s, the bolus is fully emptied, esophageal muscle fibers are relaxed, and the esophagus restores its resting configuration. We remark that the longitudinal shortening actually generates negative circumferential stress. This is illustrated in Fig.~\ref{fig_case1_sigma_uz} as a very thin blue strip in the LM layer (i.e. the outermost layer). The negative circumferential deviatoric stress in the LM layer is due to the Poisson effect. As when the LM layer shortens axially, it expands radially and circumferentially. This is also consistent with previous studies, which show that adding LM shortening will decrease the squeezing pressure~\cite{kou2015muscle,yassi2009modeling}. Fig.~\ref{fig_case1_pressure} shows the pressure field. Consistent with the clinical experiments, a peak luminal pressure is generated from the muscle contraction. And the peak luminal pressure always follows the tail of the bolus.  Notice that pressure in the structure domain (i.e. the esophageal wall) is only a Lagrange multiplier to enforce the incompressibility in the Eulerian description. Its value depends on how the elastic force is computed when the deformation of the elastic structure is given. Therefore the pressure in the structure domain obtained from the IB-FE-based model is not comparable to that obtained from our previous IB-fiber-based model. Fig.~\ref{fig_case1_kinematics} shows detailed kinematic information. Consistent with our previous findings, a typical esophageal segment passes four states during bolus transport: (a) a resting state; (b) a dilated state to accommodate the incoming bolus; (c) a contracting state as a result of the incoming muscle activation wave; (d) a relaxed state after the activation wave passes. It can be seen that the mucosal layer undergoes the largest deformation during bolus transport. This implies that the compliance of mucosal layer is important for the success of bolus transport. Fig.~\ref{fig_case1_csa_p} shows the geometrical information and pressure information along the axial direction at time = 1.2 s. Consistent with the experimental observation based on the ultrasound and concurrent manometry, a peak pressure overlaps with a peak muscle CSA behind the bolus tail. This indicates the synchrony between CM contraction and LM shortening during bolus transport. While asychrnony might indicate neural disorders that might cause motility disorder~\cite{Mittal2006,Jung04}.  

\subsubsection{Case 2: Esophageal transport using the exponential material model}
\label{sec2_exp_model}
\begin{table}[ht]
 \caption{Model parameters of the exponential model (i.e. Section~\ref{mod2_exp} ) }
 \centering
\begin{tabular}{l | l  l | l  l }
 \hline \hline
 Material type & Material parameters  \\ [1ex]
 \hline
 Mucosa & $C_0$(KPa)& 4.0e-2 & $C_1$(KPa) & 4.0e-1 \\
	& $C_2, C_3$(KPa) & 4.98e-3 & $k_2,k_3$ & 9.73e-2 \\
	&$\alpha_2^{\tmo}$(Deg.)& 48.31& $\alpha_3^{\tmo}$(Deg.)& 131.69 \\
	& $\lambda_2^{\tmo},\lambda_3^{\tmo} $& 5.0 \\
 
 \hline
 CM     & $C_4$(KPa)      & 2.17   & $k_4$ & 0.532 \\
	& $C_5 $(KPa) & 3.09     & $k_5$ & 0.532\\ 
	& $\alpha^{\tcm}$(Deg.) &0 \\
 \hline
 LM     & $C_6$(KPa)      & 2.17   & $k_6$ & 0.532 \\
	& $C_7 $(KPa) & 3.40     & $k_7$ & 0.899 \\ 
	& $\alpha^{\tlm}$(Deg.) &90 \\
 \hline
 \end{tabular}
 \label{tab_exponential_model_parameters}
\end{table}

\begin{figure}[ht] 
 \centering
 \includegraphics[scale = 0.5]{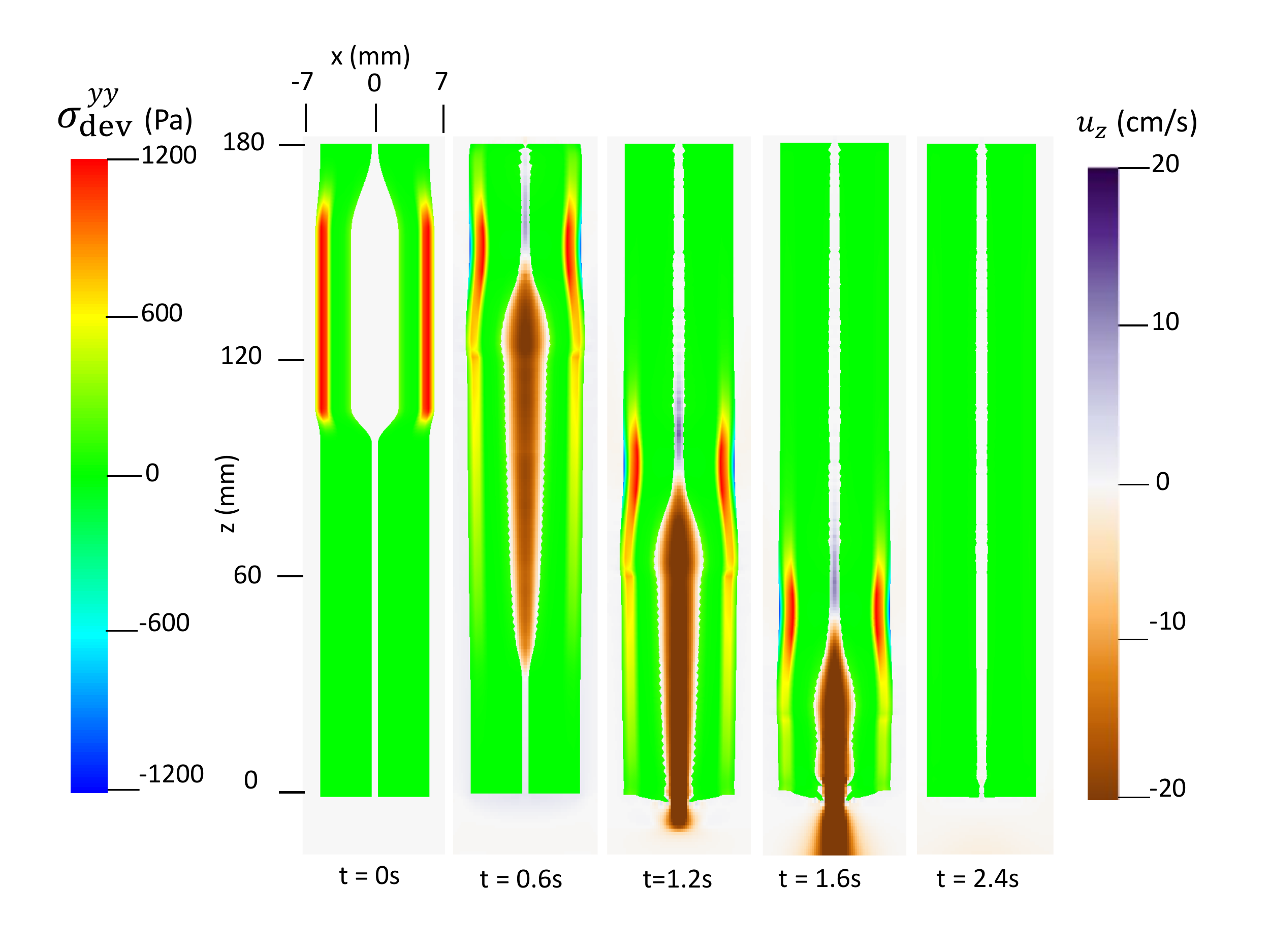} 
 \caption{Axial velocity of the bolus, $u_z$, and the yy-compoent of the deviatoric stress of the esophageal wall, $\vec{\sigma}_{dev}^{yy}$, in the plane y = 0 at different times for Case 2 in Section~\ref{sec2_exp_model}.}
 \label{fig_case2_sigma_uz}
 \end{figure}

\begin{figure}[ht] 
 \centering
 \includegraphics[scale = 0.5]{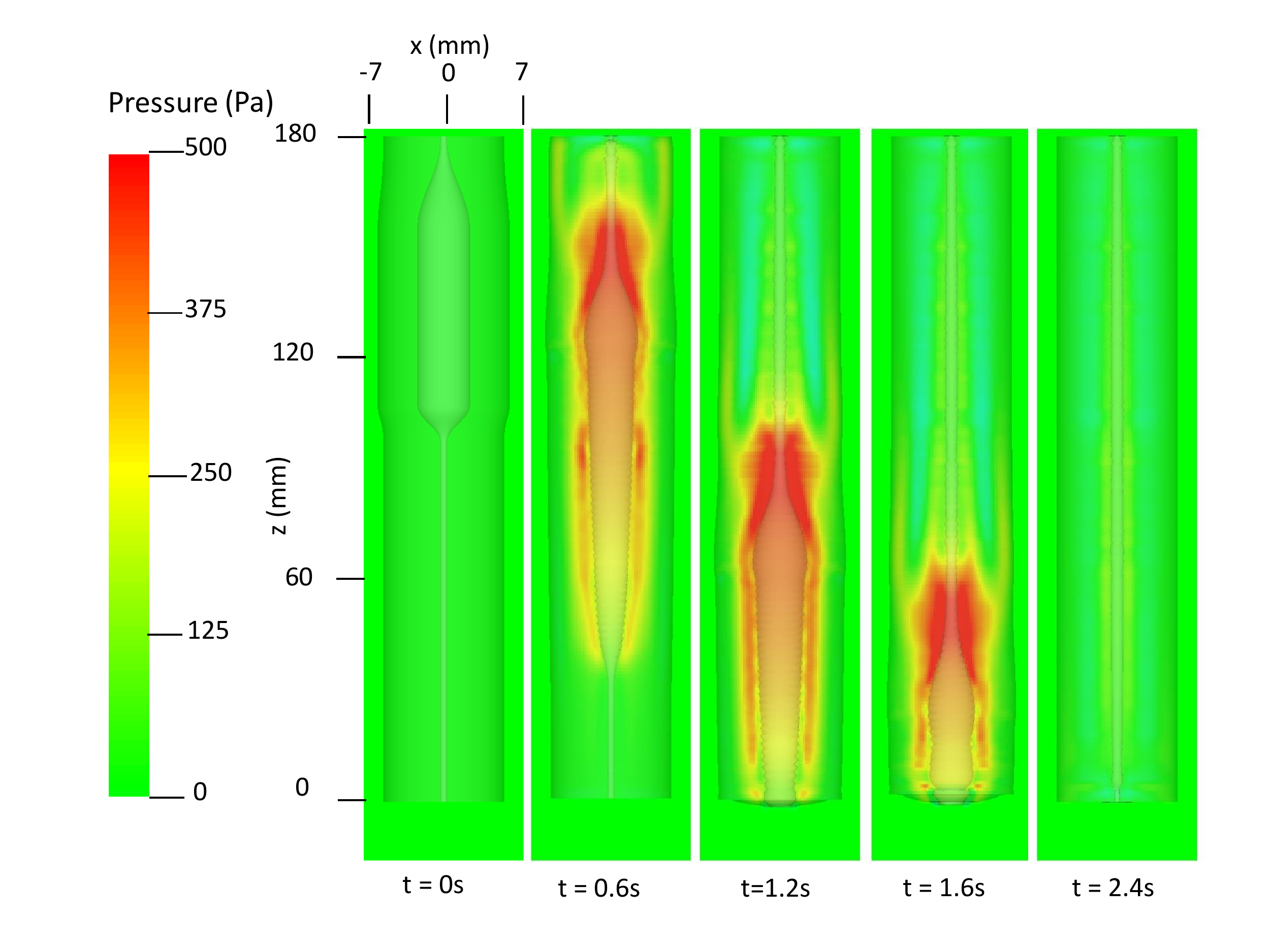} 
 \caption{Pressure field in the plane y = 0 at different times for Case 2 in Section~\ref{sec2_exp_model}.}
 \label{fig_case2_pressure}
 \end{figure}

\begin{figure}[ht] 
 \centering
 \includegraphics[scale = 0.28]{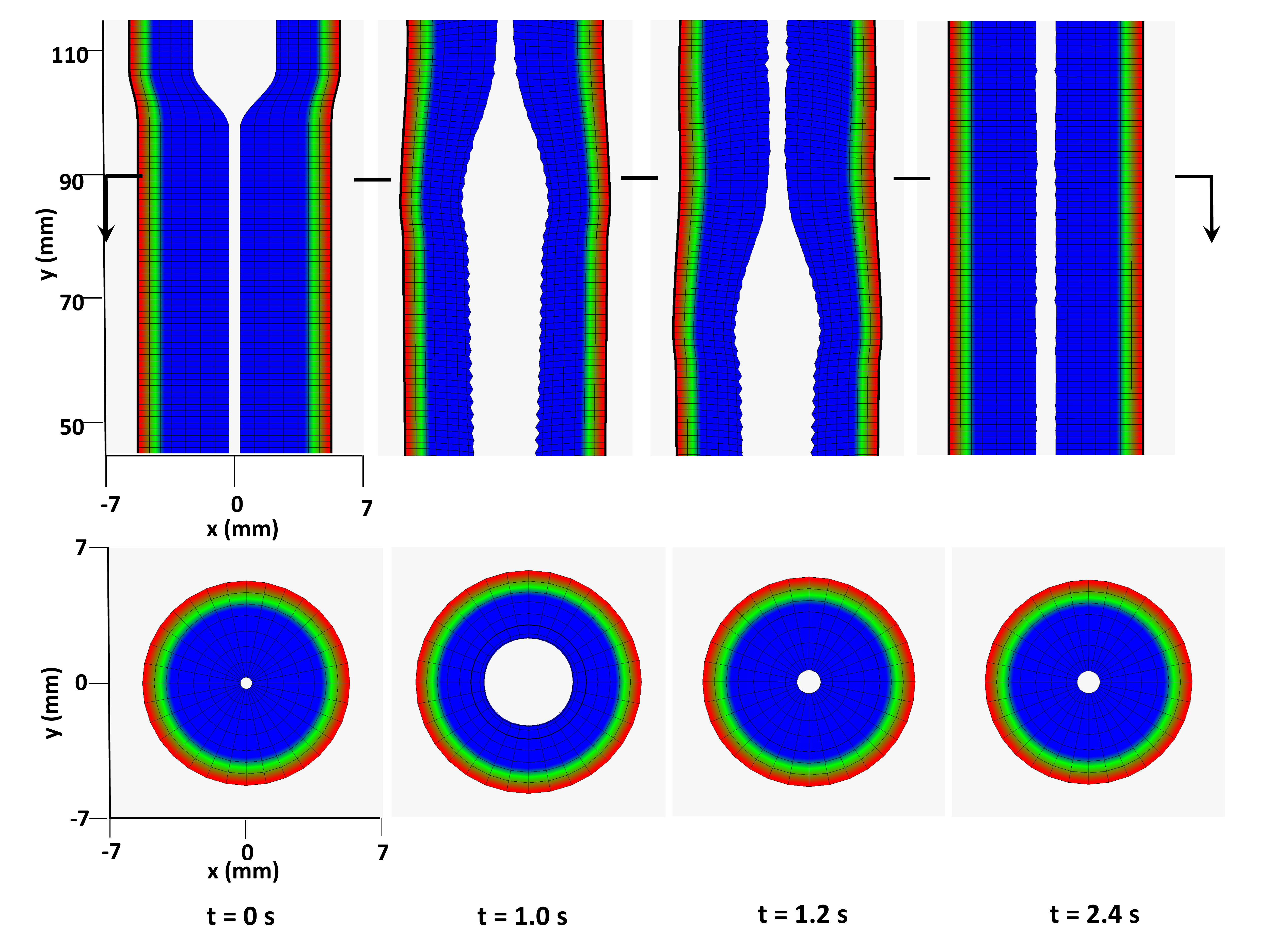} 
 \caption{Kinematics of the esophageal layers at four different stages: at rest (t = 0 s); at dilation (t = 1.0 s); at contraction (t = 1.2 s); and at relaxation (t = 2.4 s) for Case 2 in Section~\ref{sec2_exp_model}. The three layers included in the model, from the inside to the outside, are the mucosa, CM, and LM layers, respectively. (Upper) Side view of a section of the esophagus within the box: (-7 mm, 7 mm) x (-0.2 mm, 0.2 mm) x (45 mm, 115 mm); (Lower) top view of a section of the esophagus within the box: (-7 mm, 7 mm) x (-7 mm, 7 mm) x (89.5 mm, 90.5 mm). }
 \label{fig_case2_kinematics}
 \end{figure}

\begin{figure}[ht] 
 \centering
 \includegraphics[scale = 0.4]{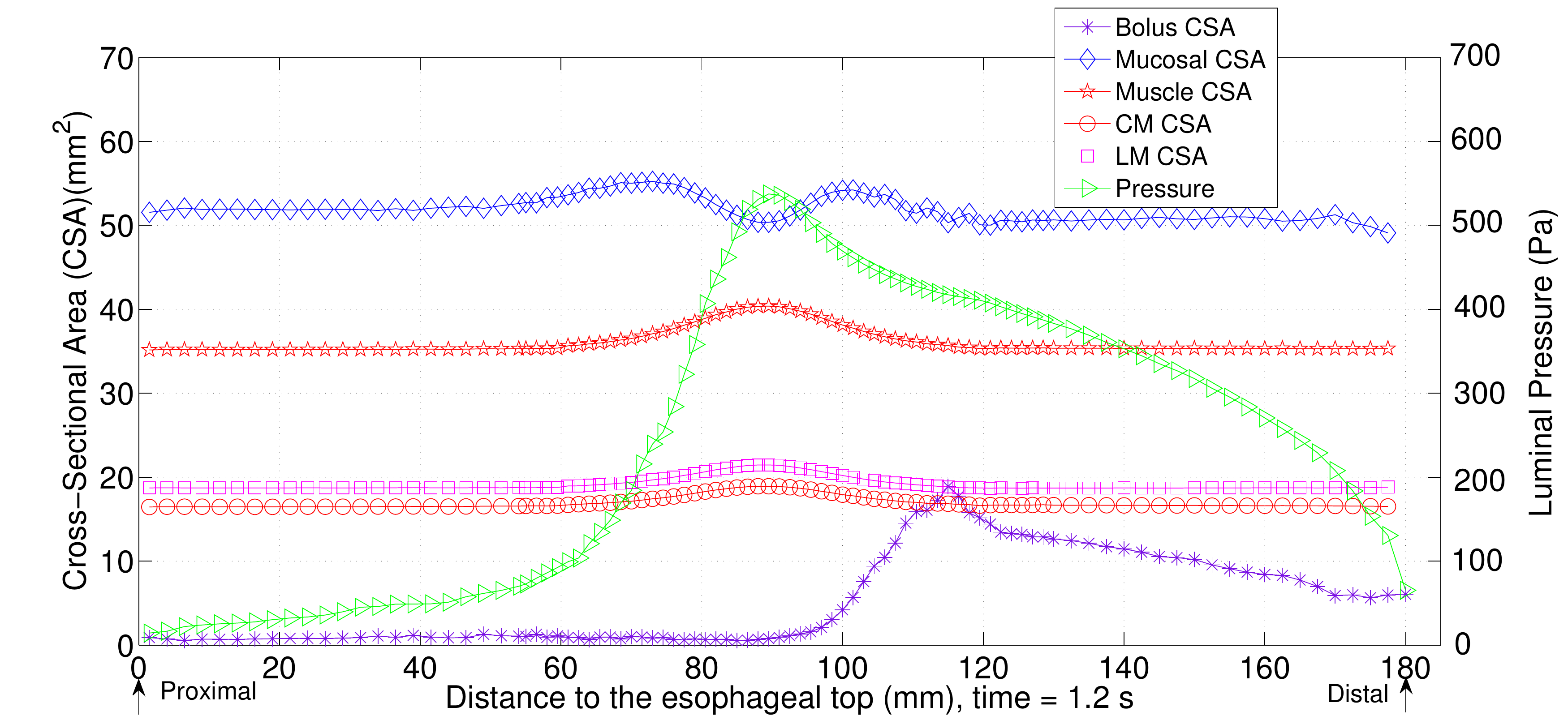} 
 \caption{The cross-sectional area (CSA) of the bolus and the esophageal components, and the lumen pressure along its central line: x = 0, y = 0, at t = 1.2 s for Case 2 in Section~\ref{sec2_exp_model}.}
 \label{fig_case2_csa_p}
 \end{figure}

Here we present our second case in which we use the exponential model (i.e. Section~\ref{mod2_exp}) for the material property of the esophagus. The material parameters are listed in Table~\ref{tab_exponential_model_parameters}. Those parameters are based on Natali et al.~\cite{Natali2009}. We adjust the moduli of esophageal layers based on our previous studies and extensive empirical tests. This is because the original parameters from in-vitro tests yield much stiff esophageal muscle and mucosa. The muscle activation model is the same as Case 1, except that we adjust the modulus of muscle fibers when they are activated. This is because the stress is an exponential function of the stretch ratio in this case. Thus, the same amount of shortening as Case 1 will yield a very high active stress. In this case, we reduce $C_5$ (or $C_7$) to one fourth of the original value in Table~\ref{tab_exponential_model_parameters}, when a CM (or LM) fiber is in the activated state. Simulation results are shown in Figs.~\ref{fig_case2_sigma_uz}-\ref{fig_case2_csa_p}. Fig.~\ref{fig_case2_sigma_uz} shows the stress and axial velocity at different times. Similarly, it illustrates that bolus transport is driven by the muscle contraction. Compared with Case 1, it shows much higher wall stress (1200 Pa vs. 500 Pa) with a longer bolus region. In particular, at t=1.2 s, the bolus begins to be emptied in this case, whereas in Case 1, the entire bolus is still confined in the esophageal body at t=1.2 s.  Fig.~\ref{fig_case2_pressure} shows the pressure field. A peak luminal pressure is located at the bolus tail, similar to Case 1. However, the pressure gradient inside the bolus seems more significant in this case. Fig.~\ref{fig_case2_kinematics} shows the detailed kinematic information. A similar four distinctive states exist for a typical segment during the transport. Fig.~\ref{fig_case2_csa_p} shows the geometrical information and pressure information along the axial direction. The luminal pressure peak overlaps with the muscle CSA peak, indicating the synchrony between CM contraction and LM shortening. Compared with Case 1, the deformation of all esophageal layers in the contracted segment seems to be less significant. However, the peak pressure and the intra-bolus pressure gradient is much higher. It seems that active contraction of an exponential fiber likely generates a higher squeezing effect. 

\subsubsection{Case 3: Esophageal transport including a more realistic and complex muscle fiber architecture}
\label{sec3_helical_fiber}
\begin{figure}[ht] 
 \centering
 \includegraphics[scale = 0.4]{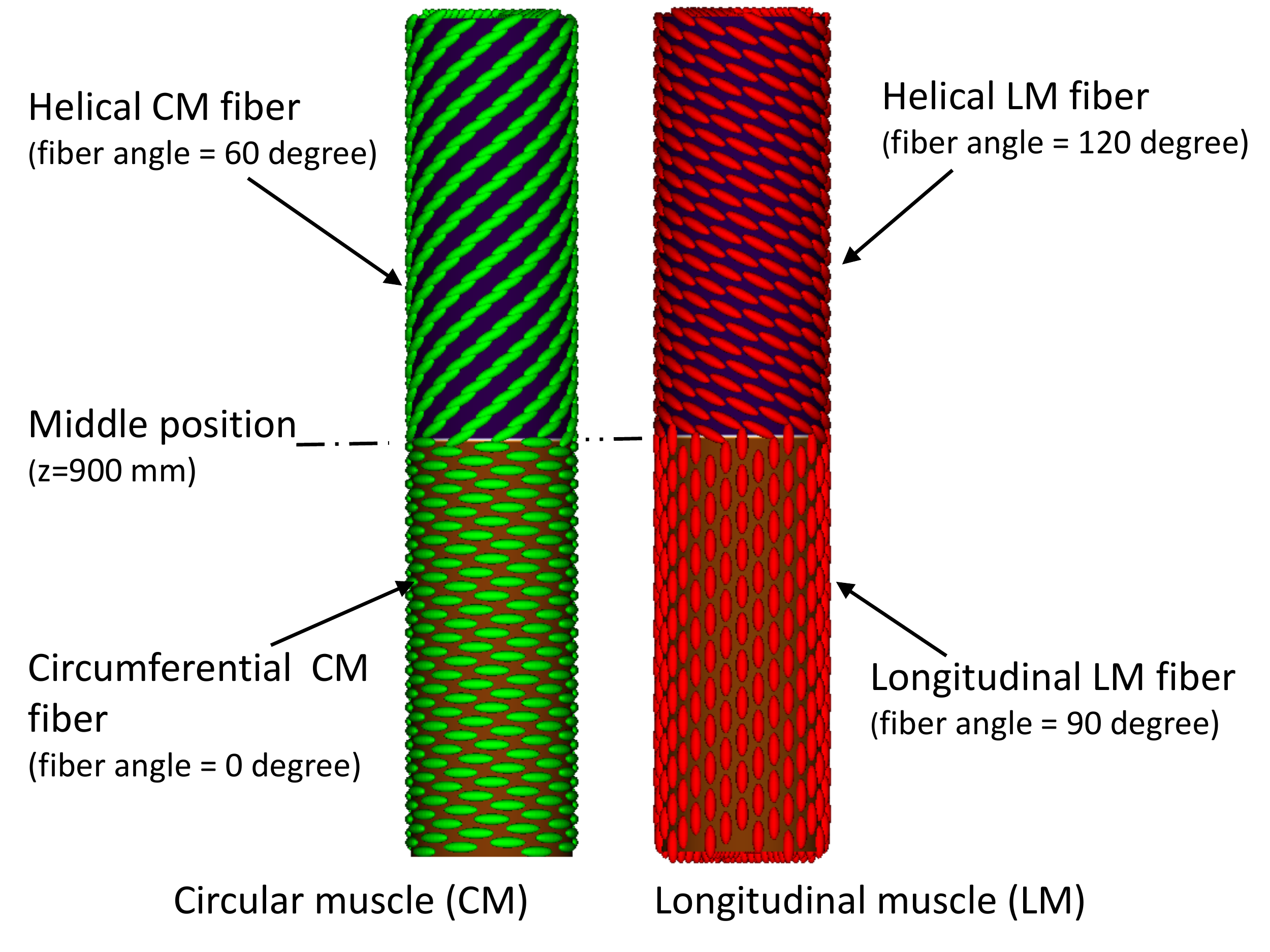} 
 \caption{Illustration of the fiber architecture in the CM layer (left) and LM layer (right) in Case 3.}
 \label{fig_case3_fibers}
 \end{figure}

We remark that compared with the previous fiber-based model, the continuum-based model in the current IB-FE framework permits us to handle more realistic and complex behavior of the esophageal tissue. We demonstrate this in our third case. Specifically, experiments show that the muscle fiber architecture of the esophagus could be very complicated. In the proximal (i.e. upper) esophageal body, the muscle fibers in both CM and LM layers are helically aligned, with helix angle up to 60-70 degree. In the distal (i.e. lower) esophageal body, the muscle fibers align themselves into distinct inner circular and outer longitudinal muscle layers~\cite{gilbert2008resolving}. To study the bio-physical consequence of this unique muscle fiber architecture, we construct a case in which the fiber orientation varies spatially in a similar way. In particular, for the upper half esophageal body, we model the muscle fibers in both CM layer and LM layer as helical fibers. The fiber angels in CM layer and LM layer are 60 and 120 degree, respectively (i.e. $\alpha_{\tcm} = 60, \alpha_{\tlm} = 120$). For the lower half esophageal body, we model the muscle fibers in CM layer and LM layer as circumferential and axial fibers, respectively (i.e. $\alpha_{\tcm} = 0, \alpha_{\tlm} = 90$). An illustration is shown in Fig.~\ref{fig_case3_fibers}.

\begin{figure}[ht] 
 \centering
 \includegraphics[scale = 0.5]{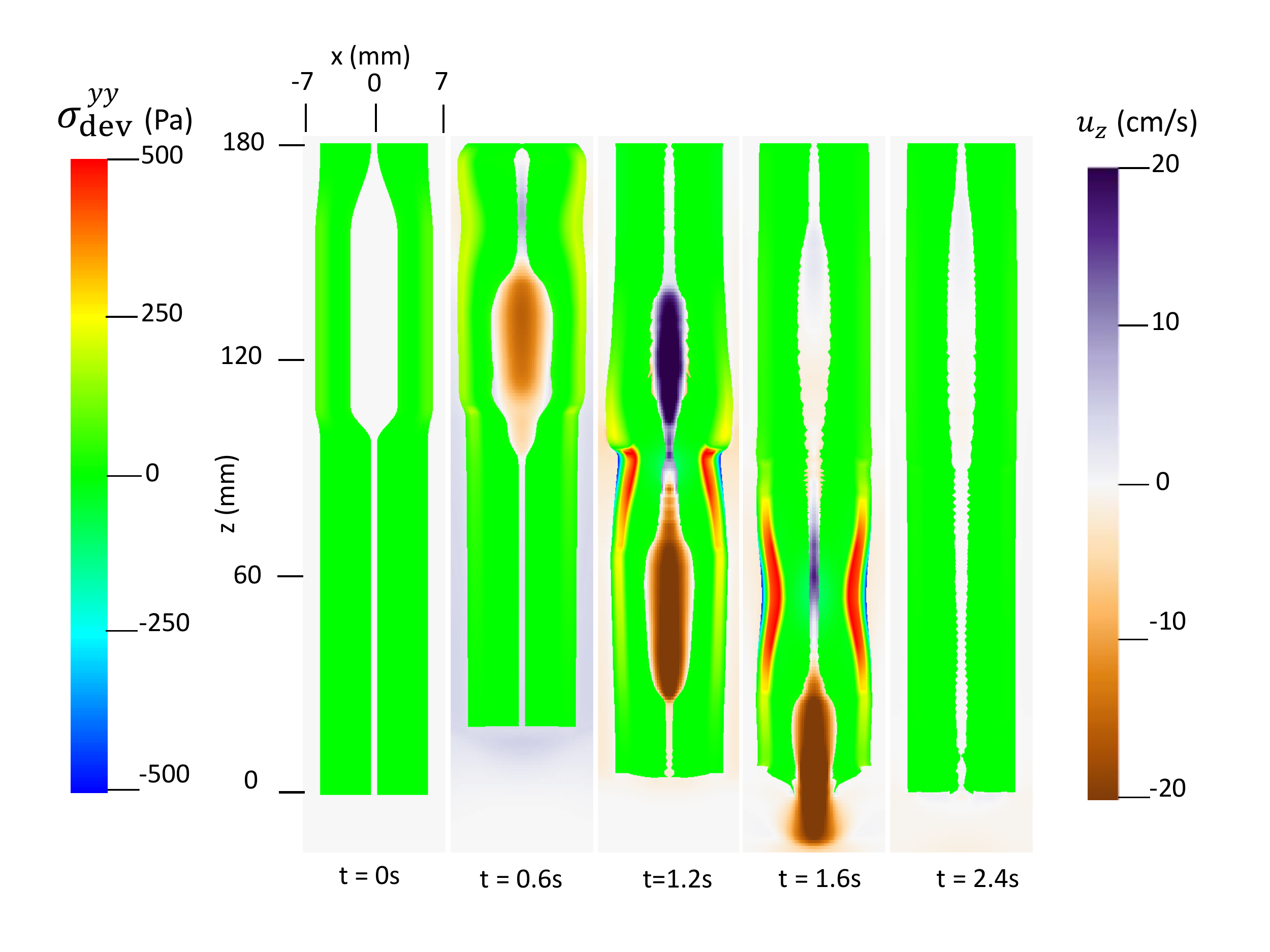} 
 \caption{Axial velocity of the bolus, $u_z$, and the yy-compoent of the deviatoric stress of the esophageal wall, $\vec{\sigma}_{dev}^{yy}$, in the plane y = 0 at different times for Case 3 in Section~\ref{sec3_helical_fiber}.}
 \label{fig_case3_sigma_uz}
 \end{figure}

\begin{figure}[ht] 
 \centering
 \includegraphics[scale = 0.5]{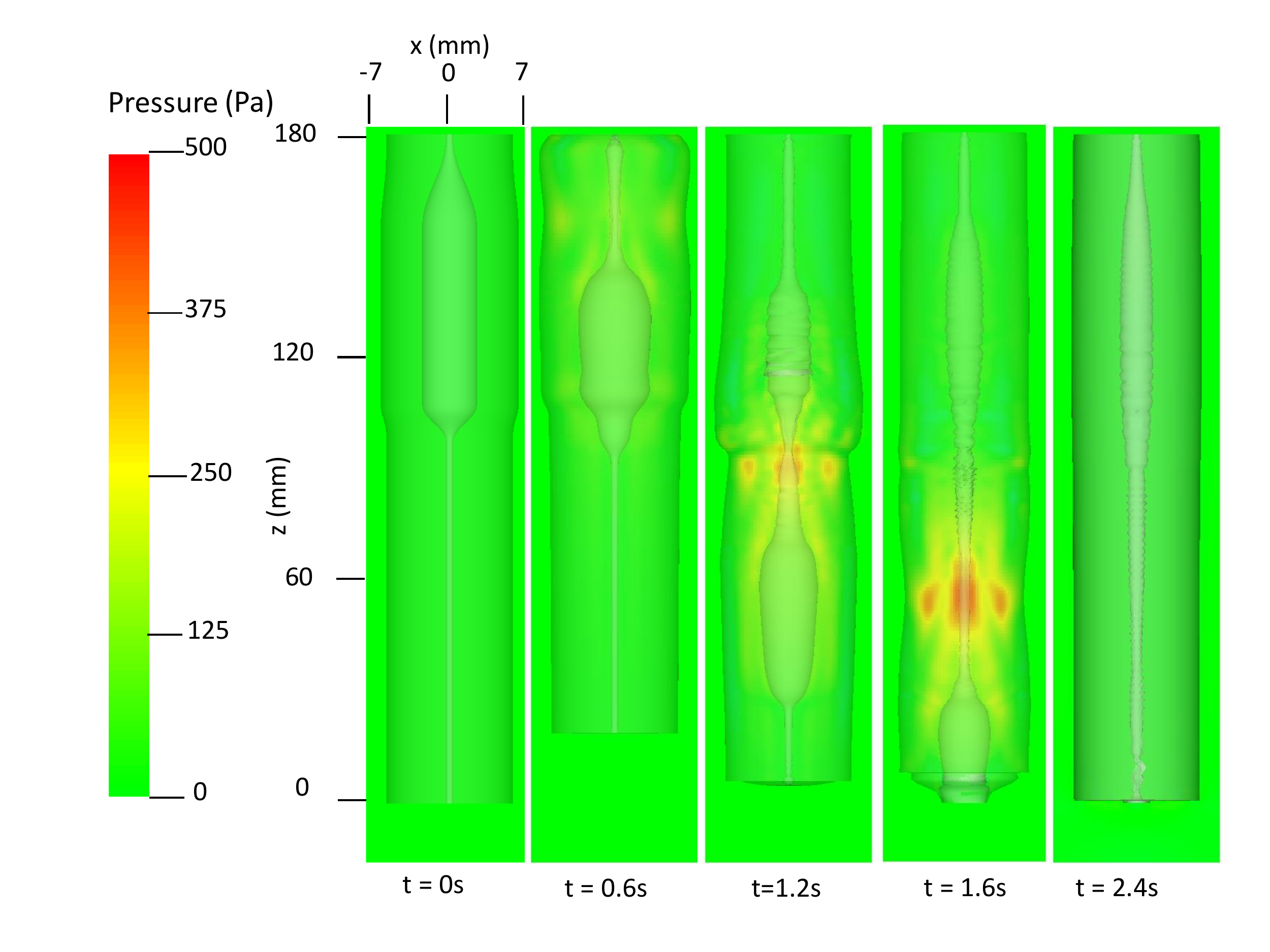} 
 \caption{Pressure field in the plane y = 0 at different times for Case 3 in Section~\ref{sec3_helical_fiber}.}
 \label{fig_case3_pressure}
 \end{figure}

In this case, we use the bi-linear model (i.e. Section~\ref{mod1_bilinear}) to describe the esophagus material property. The material parameters are the same as Case 1 (i.e. Table~\ref{tab_bilinear_model_parameters}) except that we use a spatially-varying fiber angles. The muscle activation model is also the same as Case 1. Results are shown in Figs.~\ref{fig_case3_sigma_uz}-\ref{fig_case3_pressure}. Fig.~\ref{fig_case3_sigma_uz} shows the stress and axial velocity at different times. It can be seen that the contraction of helical muscle fibers generates lower circumferential stress and a lower transport velocity than the contraction of axially-circumferentially muscle fibers. Compared with Case 1 and 2, Case 3 with a upper helical fiber architecture causes a more pronounced upward displacement of the esophagus. The jump of muscle fiber orientation also causes a jump of wall stress near the middle of the esophagus. This is likely responsible for a certain amount of the bolus retention observed near the middle region. The pressure field shown in Fig.~\ref{fig_case3_pressure} also implies that upper helical muscle fiber contraction does not generate as a high squeezing pressure as the lower part. However, we remark that Case 3 with a more realistic (i.e. upper helical and lower axial-circumferential) fiber architecture replicates a pressure pattern that is consistently observed in clinical experiments~\cite{ghosh2006physiology}. Clinical manometry on normal people shows that the spatial-temporal profile of the luminal pressure always has a pressure trough within two high pressure waves. And the lower high pressure wave is higher than the upper high pressure wave. Clinically, the pressure trough is referred to as the \textit{pressure transition zone} and considered to be a important characteristics of the normal physiology~\cite{ghosh2006physiology}. The pressure transition zone is traditionally hypothesized to be caused by multiple muscle contraction waves. However, our preliminary results suggest the spatially-varying fiber architecture might play a role in causing the pressure transition zone. Our future study will further investigate roles of the fiber architecture as well as multiple muscle contraction waves in esophageal transport. Related results will be reported separately. 

\begin{figure}[ht] 
 \centering
 \includegraphics[scale = 0.25]{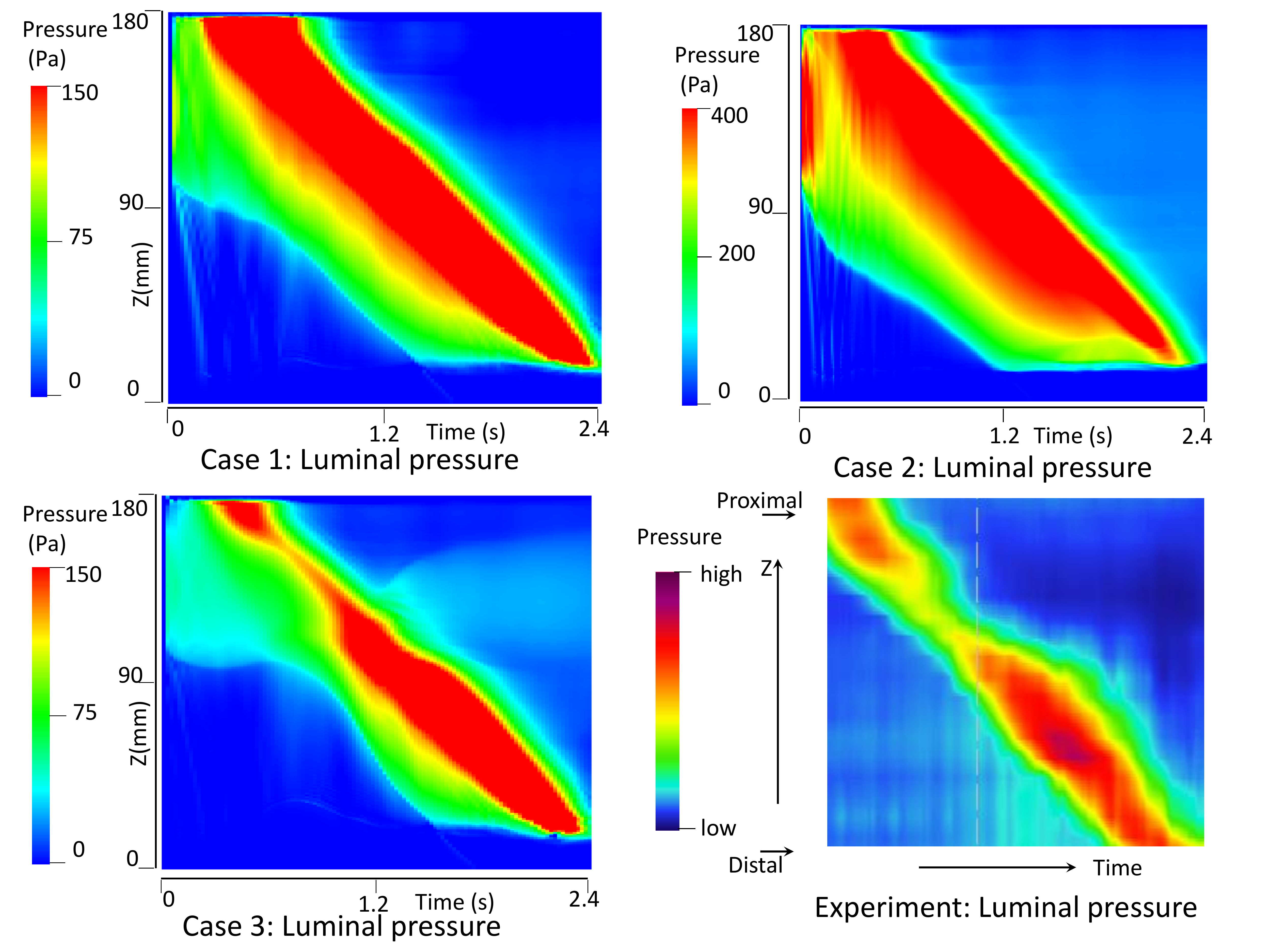} 
 \caption{Temporal-spatial profile of the luminal pressure (i.e. the pressure at $(x=0,y=0,z,t)$) obtained from Case 1 (top left), Case 2 (top right), Case 3 (bottom left), and a clinical test on a normal people (bottom right).}
 \label{fig_pressure_compare}
 \end{figure} 

\section{Conclusions}
\label{part5_conclusion}
In this work, we extend our previous IB-fiber-based model on esophageal transport to an continuum-based IB-FE model. 
We introduce an anisotropic adaptive interaction quadrature rule to handle Lagrangian-Eulerian interactions more efficiently. This adaptive rule not only helps to avoid the leakage, but also helps to reduce the computational cost. For the material model, we extend previous fiber-based material model to a continuum-based fiber-reinforced material model. The new material model can handle non-linear elasticity and fiber-matrix interactions, and thus permits us to consider more realistic material behavior of biological tissues.

To validate our methodology, we first study a case in which a three-dimensional short tube is dilated. We compare results with those obtained based on the implicit FE method. Both the pressure-displacement relationship and the stress distribution matches very well. We remark that in our IB-FE case, the three-dimensional tube undergoes a very large deformation and the Lagrangian mesh becomes much coarser than the fluid mesh. To validate the performance of the method in handling fiber-matrix material models, we perform a second verification study on dilating a long fiber-reinforced tube. We study various cases with different fiber angles, and conduct comparisons between the computational results and an analytic solution. The errors in most of the cases are less than one percent, with the largest error below 4 percent. 

We then move to our main application: esophageal transport. We adopt two fiber-reinforced models on the esophageal tissue from the literature: the bi-linear model and exponential model. We propose a simplified model on muscle contraction that is similar to our previous fiber-based work. We first show two cases. Case 1 adopts the bi-linear model to describe the esophageal material property. Case 2 adopts the exponential model. Circumferential wall stress for each case is analyzed, which is unavailable for the previous fiber-based model. The stress distribution shows clearly that the contractile stress comes from circular muscle contraction not the longitudinal muscle shortening. Information on the axial velocity, luminal pressure and kinematic information is presented. They are consistent with the observation from our previous fiber-based model. We remark that one advantage of the continuum-based model over the traditional fiber-based model is its capability to handle more realistic and complicated material behavior. This is demonstrated in our third case, in which we include a spatially-varying muscle fiber architecture based on experiments. We find that this unique muscle fiber architecture could generate an interesting luminal pressure pattern that is observed clinically. The spatial-temporal luminal pressure profile has a pressure trough, clinically called as the pressure transition zone. This preliminary study suggests the muscle fiber architecture is likely to be responsible for the pressure transition zone. Future detailed investigation through case studies is recommended.

\section*{Acknowledgments} 

The support of grant R01 DK079902 (J.E.P.) and R01 DK056033 (P.J.K.) from the National Institutes of Health, USA is gratefully acknowledged. B.E.G.~acknowledges research support from the  National Science Foundation (NSF award ACI 1450327).

\section*{Appendix A: IB-FE governing equations} \label{appendix_IBFE}

The idea of the immersed boundary method is to separate the ``fluid-like'' components in the governing equations of the structure domain. Therefore, we derive another form of eq. \eqref{eqn_momentum_s1} as below,
\begin{align}
&\rho^{\tf}\left(\D{\u^{\ts}}{t}(\x,t) + \u^{\ts}(\x,t) \cdot \grad \u^{\ts}(\x,t) \right) -\div \vsigma^{ \tilde{\tf}}, \nonumber \\
    &= \div \Delta \vsigma - \Delta \rho  \left(\D{\u^{\ts}}{t} + \u^{\ts} \cdot \grad \u^{\ts} \right), \nonumber \\
    &= \vec{f}^{\ts},\label{eqn_momentum_s2} 
\end{align}
where $\vsigma^{ \tilde{\tf}}$ is the ``fluid-like'' stress that takes the same constitutive law as the fluid stress, $\vsigma^{\tf}$. $\Delta \vsigma =\vsigma^{\ts} -\vsigma^{ \tilde{\tf}} $; $\Delta \rho =\rho^{\ts} -\rho^{\tf} $. $\vec{f}^{\ts}$ is introduced to denote all the right-hand side of eq. \eqref{eqn_momentum_s2}.

At the fluid-structure interface, eq. \eqref{eqn_momentum_fs1} can also be written as
\begin{align}
\vsigma^{\tf} \cdot \vec{n} - \vsigma^{ \tilde{\tf}} \cdot \vec{n} = \Delta \vsigma \cdot \vec{n}, \label{eqn_momentum_fs2} 
\end{align}
where $\vsigma^{\tf}$ is the fluid stress in the fluid side, and $\vsigma^{ \tilde{\tf}}$ is the ``fluid-like'' stress in the solid side.

We introduce a global velocity field, $\u(\x,t)$, such that $\u(x,t)|_{\Omega^\tf(t)} = \u^{\tf}(x,t)$, and $\u(x,t)|_{\Omega^\ts(t)} = \u^{\ts}(x,t)$.  $\u(\x,t)$ is continuous based on eq. \eqref{eqn_continuity_fs1}. We consider the fluid as an incompressible Navier-Stokes fluid, then
\begin{align}
\vsigma^{\tf} &= -p \vec{I} + \mu [\grad \u + (\grad \u)^T] \quad \text{in } \Omega^\tf(t), \\
\vsigma^{ \tilde{\tf}} &= -p \vec{I} + \mu [\grad \u + (\grad \u)^T] \quad \text{in } \Omega^\ts(t),
\end{align}
where $p$ is the pressure to enforce the incompressibility condition.

Similarly, we introduce a global fluid source $q(\x,t)$, such that, $q(\x,t)|_{\Omega^\tf(t)} =q^{\tf} (\x,t) $, and $q(\x,t)|_{\Omega^\ts(t)} = 0$. Since $q^{\tf} (\x,t)$ is specified, we restrict it to vanish at the fluid-structure interface. Therefore, $q(\x,t)$ is continuous across the fluid-structure interface. Then we obtain new governing equations as below.

\textit{In the entire domain, $\Omega$}
\begin{align}
\rho^{\tf}\left(\D{\u}{t}(\x,t) + \u(\x,t) \cdot \grad \u(\x,t) \right) &= -\grad p(\x,t) + \mu \lap \u(\x,t) + \vec{f}^{\ts}|_{\Omega^\ts(t)}, \label{eqn_momentum_whole1} \\
\div \u(\x,t) &= q (\x,t), \label{eqn_continuity_whole1} 
\end{align}
where $\vec{f}^{\ts}|_{\Omega^\ts(t)}$ is only non-zero in the structure domain, $\Omega^\ts(t)$. 

\textit{At the fluid-structure interface, ${\partial \Omega^\ts(t)}$}
\begin{align}
[|\vsigma^{\tf}|] \cdot \vec{n} &= \vsigma^{\tf} \cdot \vec{n} - \vsigma^{ \tilde{\tf}} \cdot \vec{n} = \Delta \vsigma \cdot \vec{n}, \label{eqn_stress_jump1} \\
[|\u |] &=  \u^{\tf}  - \u^{\ts} = 0,  \label{eqn_continuity_jump1} \\
\end{align}
where $[|\cdot |]$ denotes a jump in a variable across the interface, i.e. the value on the fluid side minus the value on the structure side. 

\textit{In the structure domain, $\Omega^\ts(t)$}
\begin{align}
\vec{f}^{\ts}= \div \Delta \vsigma - \Delta \rho  \left(\D{\u^{\ts}}{t} + \u^{\ts} \cdot \grad \u^{\ts} \right ).
\end{align}

Eqs. \eqref{eqn_momentum_whole1} and \eqref{eqn_stress_jump1} show that the structure influences the fluid system through two forcing terms: $\vec{f}^{\ts}$ in the solid domain, and $\Delta \vsigma \cdot \vec{n}$ at the interface. In the conventional IB method, the interface condition is not directly enforced. Instead, an delta function is used to \textit{spread} the forcing terms from the structure domain to the entire domain. Hence, the force spreading is required to include the two forcing terms. Utilizing the delta function, we obtain the immersed boundary formulation as below. We refer to this formulation as a formulation with the update Lagrangian description, as the structure is described in the current configuration.

\textit{In the entire domain, $\Omega$}
\begin{align}
\rho^{\tf} \left(\D{\u}{t}(\x,t) + \u(\x,t) \cdot \grad \u(\x,t) \right) & = \div \vec{\sigma}^{\tf} + \bar{\vec{f}}^{\ts} \nonumber \\
              & =\grad p(\x,t) + \mu \lap \u(\x,t) +  \bar{\vec{f}}^{\ts},  \label{eqn_momentum_whole2} \\
              \div \u(\x,t) &= q(\x,t), \label{eqn_continuity_whole2} \\
\bar{\vec{f}}^{\ts} &= \int_{\Omega^\ts(t)} \vec{f}^{\ts}  \delta(\x - \vec{\chi}(\s,t)) d\vec{\chi}(\s,t) \nonumber \\ 
 &- \int_{\partial \Omega^\ts(t)} \Delta \vsigma \cdot \vec{n}  \delta(\x - \vec{\chi}(\s,t)) da(\vec{\chi}(\s,t)), \label{eqn_eularfs2}  
\end{align}
where $\delta(\x)$ is the $d$-dimensional delta function.

\textit{In the structure domain, $\Omega^\ts(t)$}
\begin{align}
\vec{f}^{\ts}= \div \Delta \vsigma - \Delta \rho  \left(\D{\u^{\ts}}{t} + \u^{\ts} \cdot \grad \u^{\ts} \right ). \label{eqn_eulars2} 
\end{align}

Notice that eqs. \eqref{eqn_momentum_whole1} and \eqref{eqn_stress_jump1} are implied by eqs. \eqref{eqn_momentum_whole2} and  \eqref{eqn_eularfs2}. This can be shown as below.

\textit{For any $\x \in \Omega\setminus{\partial}\Omega^\ts(t)$}, the second term in eq. \eqref{eqn_eularfs2} drops, and eqs. \eqref{eqn_momentum_whole2} and \eqref{eqn_eularfs2} lead to eq. \eqref{eqn_momentum_whole1}. 

\begin{figure}[ht] 
 \centering
 \includegraphics[scale = 0.5]{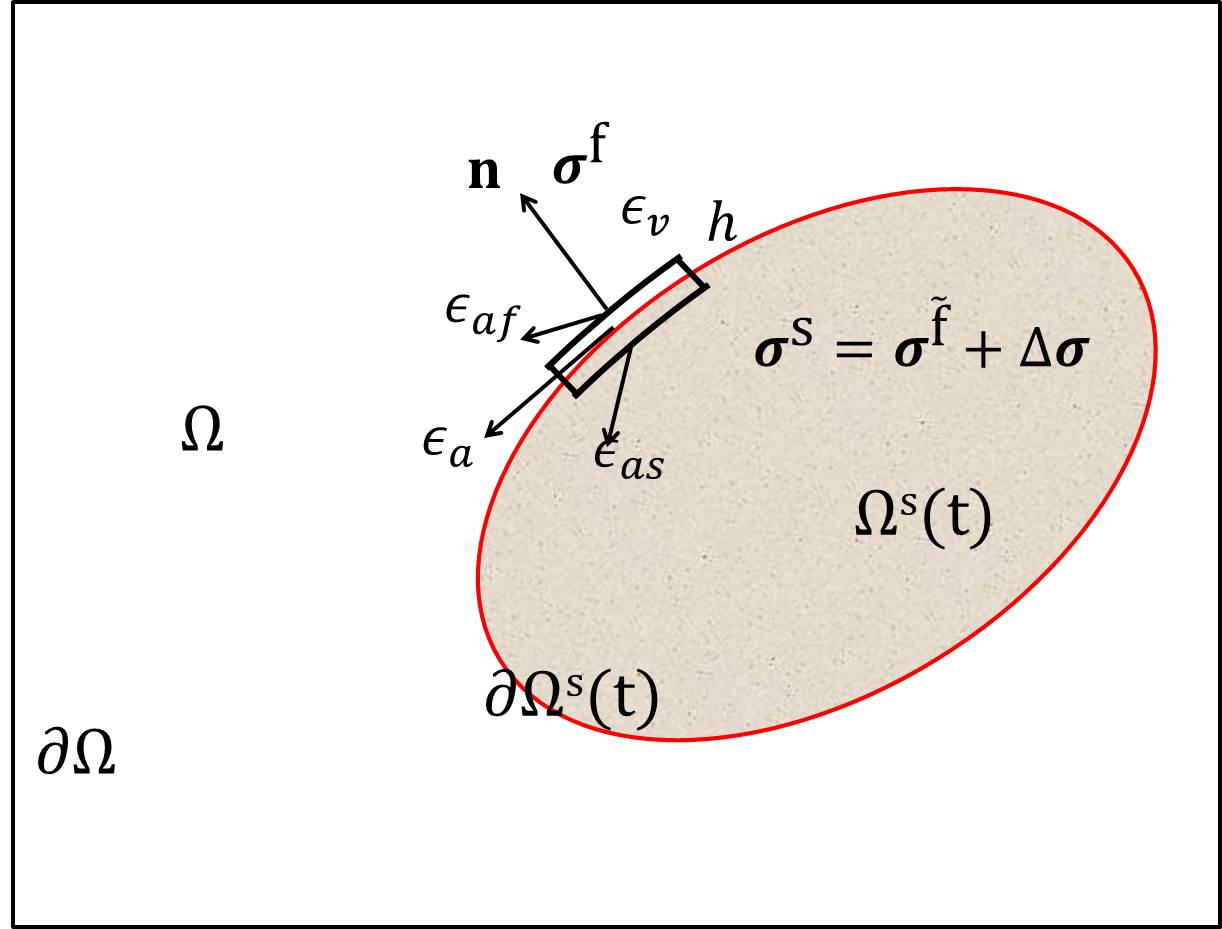} 
 \caption{Illustration of the fluid-structure system. $\Omega$ denotes the entire domain with its boundary denoted as $\partial \Omega$. $\Omega^\ts(t)$ denotes the immersed structure domain with its boundary (also the fluid-structure interface) denoted as $\partial \Omega^\ts(t)$. $\epsilon_a$ is a small portion of $\partial \Omega^\ts(t)$. Across $\epsilon_a$, a small control volume with an infinitesimal width $h$ is picked. The volume of the small control volume is denoted as $\epsilon_v$. The faces of the small control volume in the fluid domain and the structure domain are denoted as $\epsilon_{af}$ and $\epsilon_{as}$, respectively. The Cauchy stress in the fluid domain and the structure domain is denoted by $\vsigma^{\tf}$ and $\vsigma^{\ts}$, respectively. $\vsigma^{\ts}$ can be split into two parts: the fluid-like stress $\vsigma^{ \tilde{\tf}}$ and the additional stress, $\Delta \vsigma$. }
 \label{fig_formulation_fsi}
 \end{figure}

\textit{For any $\x \in \partial \Omega^\ts(t)$}, we first pick a very small surface on the interface that contains $\x$, denoted as $\epsilon_a$. Then we pick a small control volume across $\epsilon_a$ with an infinitesimal width $h$, denoted as $\epsilon_v$. $\epsilon_v$ has one face in the fluid domain, denoted as $\epsilon_{af}$ and one face in the structure domain, denoted as $\epsilon_{as}$. An illustration in the two-dimensional case is shown in Fig.~\ref{fig_formulation_fsi}. 
 
If we substitute eq. \eqref{eqn_eularfs2} into eq. \eqref{eqn_momentum_whole2} and integrate eq. \eqref{eqn_momentum_whole2} over $\epsilon_v$, we obtain
\begin{align}
  & \int_{\epsilon_v} \left( \rho^{\tf}\left(\D{\u}{t} + \u \cdot \grad \u \right) \right) d\x -\int_{\epsilon_v} \left( \int_{\Omega^\ts(t)} \vec{f}^{\ts}  \delta(\x - \vec{\chi}(\s,t)) d\vec{\chi}(\s,t) \right)d\x  \nonumber \\
  &=\int_{\epsilon_v} \left( \div \vec{\sigma}^{\tf} \right) d\x -\int_{\epsilon_v} \left( \int_{\partial \Omega^\ts(t)} \Delta \vsigma \cdot \vec{n}  \delta(\x - \vec{\chi}(\s,t)) da \right) d\x \nonumber \\
  &= \epsilon_{af} (\vec{\sigma}^{\tf} \cdot \vec{n}) -\epsilon_{as} (\vsigma^{ \tilde{\tf}} \cdot \vec{n}) - \epsilon_a (\Delta \vsigma \cdot \vec{n} ). \label{eqn_jumpStress2_1}
\end{align}
Notice that the above eq. \eqref{eqn_jumpStress2_1} holds for any arbitrarily small $h$. Thus, if we let $ h \rightarrow 0$, then the left-hand side of eq. \eqref{eqn_jumpStress2_1} goes to zero, and $\epsilon_{af} \rightarrow  \epsilon_{as} \rightarrow  \epsilon_a$. So we obtain 
\begin{align}
  \vec{\sigma}^{\tf} \cdot \vec{n} - \vsigma^{ \tilde{\tf}} \cdot \vec{n} - \Delta \vsigma \cdot \vec{n} = 0. \label{eqn_jumpStress2_2}
\end{align}
This is the same as eq. \eqref{eqn_stress_jump1}.
We remark that $\bar{\vec{f}}^{\ts}|_{\Omega^\ts(t)} \neq \vec{f}^{\ts}$. $\bar{\vec{f}^{\ts}}$ also contains the forcing information along the boundary of the structure domain, in order to satisfy the force balance across the fluid-structure interface, i.e. eq. \eqref{eqn_stress_jump1}.

$\bar{\vec{f}^{\ts}}$ depends on the material property of both the fluid and the solid. In our current work, we consider that the fluid-structure system possesses a uniform mass density $\rho $, i.e. $\rho^{\ts} =\rho^{\tf} =\rho $, and a uniform dynamic viscosity $\mu$. This simplification implies that the immersed structure is neutrally buoyant and viscoelastic rather than purely elastic. We denote the elastic stress in the current configuration, i.e. Cauchy elastic stress, as $\vec{\sigma}^{\te}$. Then $\vsigma^{\ts} = \vsigma^{ \tilde{\tf}} + \vsigma^{\te}, \Delta \vsigma = \vsigma^{\te}$. We let $\vec{f}^{\te}$ denote $\bar{\vec{f}}^{\ts}$ to be consistent with the convention, where $\vec{f}^{\te}$ is referred to as the \textit{Eulerian force density}. Then eqs. \eqref{eqn_eulars2} and \eqref{eqn_eularfs2} become
\begin{align}
\vec{f}^{\ts} &= \div \vsigma^{\te}, \\
\vec{f}^{\te} &= \bar{\vec{f}}^{\ts} = \int_{\Omega^\ts(t)} \div \vsigma^{\te}  \delta(\x - \vec{\chi}(\s,t)) d\vec{\chi}(\s,t) \nonumber \\ 
 &- \int_{\partial \Omega^\ts(t)} \vsigma^{\te} \cdot \vec{n}  \delta(\x - \vec{\chi}(\s,t)) da(\vec{\chi}(\s,t)). \label{eqn_eularf2}  
\end{align}

\section*{Appendix B: Analysis of a fiber-reinforced tube dilation problem} \label{appendix_derivation}

Here we derive eq.~\eqref{eqn_validation_case2_Pinner} for the tube dilation problem presented in 
Section~\ref{sec_tube_dilation}. We first derive the stress-strain relationship. The elastic potential of the fiber-reinforced material is given in eqs. \eqref{eqn_validation_case2_psi}-\eqref{eqn_validation_case2_psif}. Then, from eq. \eqref{eqn_elastic_model_PK1}, we can compute the first Piola-Kirchhoff stress as below,

\begin{align}
\tP &= 2 \D{\varPsim}{I_1} \tensor{F} + 2 \sum_i \D{\varPsif}{I_{\fbi}} \tensor{F}\tensor{A}_{\fbi} \nonumber \\
    &= C_1 \tensor{F} + C_2 \sum_i \left ( 1- 1/ \sqrt{I_{\fbi}} \right ) \tensor{F} \tensor{A}_{\fbi}. \label{eqn_appendix_PK1}
\end{align}
Notice that the relationship between the Cauchy stress $\vec{\sigma}$ and the first Piola-Kirchhoff stress is
\begin{equation}
 \vec{\sigma} = J^{-1} \tensor{P} \tensor{F}^{T}. \label{eqn_appendix_sigma}
\end{equation}
Substituting eq. \eqref{eqn_appendix_PK1} into eq. \eqref{eqn_appendix_sigma} and considering that our material is incompressible, we obtain the expression for the Cauchy stress as below,
\begin{equation}
 \vec{\sigma} = C_1 \tensor{F} \tensor{F}^{T} + C_2 \sum_i \left ( 1- 1/ \sqrt{I_{\fbi}} \right ) \tensor{F} \tensor{A}_{\fbi} \tensor{F}^{T}.  \label{eqn_appendix_sigma_2}
\end{equation}
Here, we include two families of fiber, with fiber angles as $\alpha$ and $180 - \alpha$, respectively. We choose the cylindrical coordinate system for both the initial configuration and the deformed configuration, which are labeled as $(R, \Theta, Z)$, and $(r,\theta, z)$ respectively. Then the orientation of the two families of fibers in the initial configuration is $\vec{a}_1 = (0 \hat R , \cos \alpha \hat \Theta, \sin \alpha \hat Z)$, and $\vec{a}_2 = (0 \hat R , -\cos \alpha \hat \Theta, \sin \alpha \hat Z)$, respectively. And $\tensor{A}_{\fbi} = \vec{a}_i \otimes \vec{a}_i, (i=1,2)$.

We remark that the axially symmetric deformation exists due to the fact that fiber angles of the two families of fibers sum up to 180 degree. This can be shown by demonstrating that: if the deformation gradient $\tensor{F}$ is axially symmetric, the Cauchy stress will also be axially symmetric. Hence, let's assume $\tensor{F}$ is axially symmetric, 
\begin{equation}
 \tensor{F} = \lambda_r \hat r \otimes \hat R + \lambda_\theta \hat \theta \otimes \hat \Theta + \lambda_z \hat z \otimes \hat Z,  \label{eqn_appendix_FF}  
\end{equation}
where $\lambda_r$, $\lambda_\theta$, and $\lambda_z$ are stretch ratios along $\hat r$, $\hat \theta$, $\hat z$, respectively. With the deformation gradient, we can get several equations as below,
\begin{align}
 \tensor{F} \tensor{F}^{T} &= \lambda_r^2 \hat r \otimes \hat r + \lambda_\theta^2 \hat \theta \otimes \hat \theta + \lambda_z^2 \hat z \otimes \hat z,  \label{eqn_appendix_FFt} \\
 I_{\fb 1} &=  I_{\fb 2} = \lambda_\theta^2 \cos^2 \alpha + \lambda_z^2 \sin^2 \alpha \\
\tensor{F} \tensor{A}_{\fb 1} \tensor{F}^{T} + \tensor{F} \tensor{A}_{\fb 2} \tensor{F}^{T} &= 0 \hat r \otimes \hat r + 2 \lambda_\theta^2 \cos^2 \alpha \hat \theta \otimes \hat \theta + 2 \lambda_z^2 \sin^2 \alpha \hat z \otimes \hat z.  \label{eqn_appendix_FAF}
 \end{align}
Substituting eqs. \eqref{eqn_appendix_FFt}-\eqref{eqn_appendix_FAF} in to eq. \eqref{eqn_appendix_sigma_2}, we get,
\begin{align}
 \vec{\sigma} &= \sigma_r \hat r \otimes \hat r + \sigma_\theta \hat \theta \otimes \hat \theta + \sigma_z \hat z \otimes \hat z, \label{eqn_appendix_sigma_total} \\
 \sigma_r &= C_1 \lambda_r^2, \label{eqn_appendix_sigma_r} \\
 \sigma_\theta &=C_1 \lambda_\theta^2 + 2C_2 \left ( 1- 1/ \sqrt{\lambda_\theta^2 \cos^2 \alpha + \lambda_z^2 \sin^2 \alpha} \right ) \lambda_\theta^2 \cos^2 \alpha, \label{eqn_appendix_sigma_theta} \\
 \sigma_z &= C_1 \lambda_z^2 +2C_2 \left ( 1- 1/ \sqrt{\lambda_\theta^2 \cos^2 \alpha + \lambda_z^2 \sin^2\alpha} \right ) \lambda_z^2 \sin^2 \alpha. \label{eqn_appendix_sigma_z}  
\end{align}
The above eqs. \eqref{eqn_appendix_sigma_total}-\eqref{eqn_appendix_sigma_z} show that the Cauchy stress is indeed axially symmetric. Since the deformation is symmetric and the volume does not change, we have geometric relationships as below,
\begin{align}
\lambda_r&= \frac{r}{R};  \lambda_\theta = \frac{dr}{dR}, \label{eqn_appendix_r_dr} \\
1 &= \lambda_\theta \lambda_r \lambda_z = \frac{r}{R} \frac{dr}{dR} \lambda_z,  \label{eqn_appendix_drdR}\\
r(R_i) &= r_i; r(R_o) = r_o, \label{eqn_appendix_ri_ro}
\end{align}
where $r_i$ and $r_o$ is the deformed inner and outer radius, respectively. $R_i$ and $R_o$ is the initial inner and outer radius, respectively. Based on eqs. \eqref{eqn_appendix_drdR}-\eqref{eqn_appendix_ri_ro}, we can solve $r(R)$ and $\lambda_z$, as below,

\begin{align}
 r(R) &= \sqrt{R^2 \left ( \frac{r_o^2 -r_i^2 }{R_o^2 -R_i^2} \right ) + r_i^2 - R_i^2 \left( \frac{r_o^2 -r_i^2 }{R_o^2 -R_i^2} \right ) }, \label{eqn_appendix_rR} \\
 \lambda_z&=  \frac{R_o^2 -R_i^2}{r_o^2 -r_i^2 }. \label{eqn_appendix_lambdaz} 
\end{align}

We also have equations from the force balance along the radial direction as below,
\begin{align}
 \frac{d\sigma_r}{dr} + \frac{\sigma_r - \sigma_\theta}{r} &= 0, \label{eqn_appendix_sigma_r_theta}\\
 \sigma_r(r = r_{o}) &= 0, \label{eqn_appendix_sigma_r_outer} \\
 \sigma_r(r = r_{i}) &= - P_{inner}. \label{eqn_appendix_sigma_r_inner} 
\end{align}
Thus, $P_{inner}$ can be expressed as
\begin{equation}
P_{\text{inner}} =  \int_{r_{i}}^{r_{o}} \frac{\sigma_\theta - \sigma_r}{r} dr =\int_{R_{i}}^{R_{o}} \frac{\sigma_\theta - \sigma_r}{r} \frac{dr}{dR} dR.  \label{eqn_appendix_P_Ri_Ro}  
\end{equation}

Substituting eq. \eqref{eqn_appendix_r_dr} into eqs. \eqref{eqn_appendix_sigma_total}-\eqref{eqn_appendix_sigma_z} and eq. \eqref{eqn_appendix_P_Ri_Ro}, we get our final equation as below,

\begin{align}
 P_{\text{inner}} &=  \int_{R_{i}}^{R_{o}} \frac{C_1}{r} \left[\frac{r^2}{R^2} -\left (\frac{dr}{dR}\right)^2 \right ]\frac{dr}{dR} dR \nonumber \\
 &+ \int_{R_{i}}^{R_{o}} \frac{2C_2}{r} \left(\frac{r \cos\alpha}{R} \right)^2 \left [1 - 1/\sqrt{\left (\frac{r \cos\alpha}{R}\right)^2 + (\lambda_z \sin\alpha)^2 } \right ] \frac{dr}{dR} dR, \label{eqn_appendix_Pinner_final}
\end{align}
where $r(R)$ and $\lambda_z$ is calculated based on eqs. \eqref{eqn_appendix_rR} and \eqref{eqn_appendix_lambdaz}.


\begin{flushleft}
  \bibliography{EsoTransport}
\end{flushleft}

\end{document}